\documentclass[11pt]{article}

\usepackage{color,amsfonts,epsfig,amsmath,psfig,fullpage}

\begin{document}

\def\a{\alpha}

\newcommand{\comment}[1]{}
\newcommand{\boost}{\mathrm{boost}}

\newcommand{\C}{{\cal C}}

\newcommand{\dalpha}{{\rm d}\alpha}

\def\reals{\mathbb{R}}
\def\ones{\mathbf{1}}
\def\naturals{\mathbb{N}}

\def\eg{e.g.,\ }
\def\ie{i.e.,\ }

\def\as{a.s.\ }
\def\whp{w.h.p.}
\def\wupp{w.u.p.p.}
\def\Whp{W.h.p.}

\newcommand{\fo}[2]{\ensuremath{{F}_{#1}(n,#2)}}
\def\var{{\rm var}}
\def\ex{{\bf E}}

\newtheorem{definition}{Definition}
\newtheorem{lemma}{Lemma}
\newtheorem{claim}{Claim}
\newtheorem{theorem}{Theorem}
\newtheorem{corollary}{Corollary}
\newtheorem{conjecture}{Conjecture}

\newcommand{\ann}{{\rm ann}}
\newcommand{\qed}{{\hfill $\Box$} \vskip5mm}
\newcommand{\proofend}{\smallskip \hfill\mbox{$\Box$}\\}
\newcommand{\ra}{\rightarrow}

\newcommand{\dima}{}
\newcommand{\poor}{}

\newcommand{\e}{{\rm e}}
\newcommand{\eps}{\epsilon}
\newcommand{\beq}{\begin{equation}}
\newcommand{\eeq}{\end{equation}}
\newcommand{\bea}{\begin{eqnarray*}}
\newcommand{\eea}{\end{eqnarray*}}
\newcommand{\sub}{\scriptsize}
\newcommand{\mat}{\left(\!\!\begin{array}{cc}}
\newcommand{\rix}{\end{array}\!\!\right)}
\newcommand{\ord}{{O}}
\newcommand{\bra}{\langle}
\newcommand{\ket}{\rangle}
\newcommand{\p}{\partial}
\newcommand{\Z}{{\Bbb Z}}
\newcommand{\dm}{{\rm d}m}
\newcommand{\ds}{{\rm d}s}
\newcommand{\dt}{{\rm d}t}
\newcommand{\dv}{{\rm d}v}
\newcommand{\dx}{{\rm d}x}
\newcommand{\dy}{{\rm d}y}
\newcommand{\dw}{{\rm d}w}
\newcommand{\db}{{\rm d}b}
\newcommand{\dc}{{\rm d}c}
\newcommand{\du}{{\rm d}u}
\newcommand{\dbeta}{{\rm d}\beta}
\newcommand{\dgamma}{{\rm d}\gamma}
\newcommand{\ddelta}{{\rm d}\delta}
\def\limninf{\lim_{n \rightarrow \infty}}

\renewcommand{\d}{{\delta}}
\newcommand{\amax}{\alpha_{\max}}
\newcommand{\bmax}{\beta_{\max}}
\newcommand{\gamax}{\gamma_{\max}}
\newcommand{\gmax}{g_{\max}}
\newcommand{\za}{\vec{\zeta}}
\newcommand{\zamax}{\vec{\zeta}_{\max}}
\newcommand{\HTC}{{\rm H}}
\newcommand{\NAESAT}{{\rm N}}

\title{Random $k$-SAT: Two Moments Suffice to Cross a Sharp Threshold}

\author{
Dimitris Achlioptas \\
Microsoft Research, Redmond, Washington \\
{\tt optas@microsoft.com} \and
Cristopher Moore \\
Computer Science Department, University of New Mexico, Albuquerque \\
and the Santa Fe Institute, Santa Fe, New Mexico \\
{\tt moore@cs.unm.edu}
}

\date{}
\maketitle

\begin{abstract}
Many NP-complete constraint satisfaction problems appear to
undergo a ``phase transition'' from solubility to insolubility
when the constraint density passes through a critical threshold.
In all such cases it is easy to derive upper bounds on the
location of the threshold by showing that above a certain density
the first moment (expectation) of the number of solutions tends to
zero. {We show that in the case of certain symmetric constraints,
considering the second moment of the number of solutions yields
nearly matching lower bounds for the location of the threshold.} Specifically, 
we prove that the threshold for both random hypergraph
2-colorability (Property B) and random Not-All-Equal $k$-SAT is
$2^{k-1} \ln2 -O(1)$. As a corollary, we establish that the
threshold for random $k$-SAT is of order $\Theta(2^k)$, resolving
a long-standing open problem.
\end{abstract}

\section{Introduction}

In the early 1900s, Bernstein~\cite{bernstein} asked the following
question: given a collection of subsets of a set $V$, is there a
partition of $V$ into $V_1,V_2$ such that no subset is contained
in either $V_1$ or $V_2$? If we think of the elements of $V$ as
vertices and of each subset as a hyperedge, the question can be
rephrased as whether a given hypergraph can be 2-colored so that
no hyperedge is monochromatic. Of particular interest is the
setting where all hyperedges contain $k$ vertices, \ie $k$-uniform
hypergraphs. 
This question was popularized by Erd\H{o}s -- who dubbed it
``Property B'' in honor of Bernstein -- and  has motivated some of
the deepest advances in probabilistic combinatorics. Indeed,
determining the smallest number of hyperedges in a non-2-colorable
$k$-uniform hypergraph remains one of the most important problems
in extremal graph theory, perhaps second only to the Ramsey
problem~\cite{probmethod}.

A more modern problem, with a somewhat similar flavor, is Boolean
Satisfiability: given a CNF formula $F$, is it possible to assign
truth values to the variables of $F$ so that it evaluates to true?
Satisfiability has been the central problem of computational
complexity since 1971 when Cook~\cite{cook} proved that it is
complete for the class NP. The case where all clauses have the
same size $k$ is known as $k$-SAT and is NP-complete for all
$k\geq 3$.

For both $k$-SAT and Property B it is common to generate random
instances by selecting a corresponding structure at random.
{Indeed, random formulas and random hypergraphs have been studied
extensively in probabilistic combinatorics in the last three
decades.  While there are a number of slightly different models
for generating such structures ``uniformly at random", we will see
that results transfer readily between them.} For the sake of
concreteness,
let $F_k(n,m)$ denote a formula chosen uniformly among all
$\binom{2^k \binom{n}{k}}{m}$ \mbox{$k$-CNF} formulas on $n$
variables with $m$ clauses. Similarly, let $H_k(n,m)$ denote a
hypergraph chosen uniformly among all $\binom{\binom{n}{k}}{m}$
\mbox{$k$-uniform} hypergraphs with $n$ vertices and $m$
hyperedges. We will say that a sequence of events ${\cal E}_n$
occurs {\em with high probability} (\whp) if $\lim_{n \to \infty}
\Pr[{\cal E}_n] = 1$ and {\em with uniformly positive
probability\/} (\wupp) if $\liminf_{n \to \infty} \Pr[{\cal E}_n]
>0$. Throughout the paper, $k$ will be arbitrarily large
but fixed.

In recent years, both problems have been understood to undergo a
``phase transition'' as the ratio of constraints to variables
passes through a critical threshold. That is, for a given number
of vertices {(variables),} the probability that a random instance
has a solution drops rapidly from 1 to 0 around a critical number
of hyperedges {(clauses)}. This sharp threshold phenomenon was
discovered in the early 1990s, when several
researchers~\cite{cheeseman,MSL} performed computational
experiments on $F_3(n,m=rn)$ and found that while for $r < 4.1$
almost all formulas are satisfiable, for $r > 4.3$ almost all are
unsatisfiable. Moreover, as $n$ increases, this transition narrows
around $r \approx 4.2$. Along with similar results for other fixed
$k\geq 3$ this has led to the following popular conjecture:
\medskip

\noindent {\bf Satisfiability Threshold Conjecture:} {\em For each
$k \geq 3$, there exists a constant $r_k$ such that}
\[
\limninf \Pr[F_k(n,rn) \mbox{ is satisfiable}]=
\begin{cases}
1 & \mbox{if $r<r_k$}\\
0 & \mbox{if $r>r_k$} \enspace .
\end{cases}
\]

In the last ten years, this conjecture has become an active area
of interdisciplinary research, receiving attention in theoretical
computer science, artificial intelligence, combinatorics and, more
recently, statistical physics. Much of the work on random $k$-SAT
has focused on proving upper and lower bounds for $r_k$, both for
the smallest computationally hard case $k=3$ and for general $k$.
At this point the existence of $r_k$ has not been established for
any $k \geq 3$. Nevertheless, we will take the liberty of writing
$r_k \geq r^*$ to denote that for all $r<r^*$,  \fo{k}{rn} is
\whp\ satisfiable; analogously, we will write $r_k \leq r^*$ to
denote that for all $r>r^*$, \fo{k}{rn} is \whp\ unsatisfiable.

{As we will see, an} elementary counting argument yields $r_k \leq
2^k \ln 2$ for all $k$. {Lower bounds, on the other hand,
 have} been exclusively algorithmic: to establish $r_k \geq r^*$
ones shows that for $r<r^*$ some specific algorithm finds a
satisfying assignment with probability \mbox{that tends to 1.} We
will see that an extremely simple algorithm~\cite{mick} already
yields $r_k =\Omega(2^k/k)$. We will also see that while more
sophisticated algorithms improve this bound slightly, to date no
algorithm is known to find a satisfying truth assignment (even
\wupp) when $r = \omega(k)\times 2^k/k$, for any $\omega(k) \to
\infty$.

The threshold picture for hypergraph 2-colorability is completely
analogous: for each $k\geq 3$,  it is conjectured that there
exists a constant $c_k$ such that
\[
\limninf \Pr[H_k(n,cn) \mbox{ is 2-colorable}]=
\begin{cases}
1 & \mbox{if $c<c_k$}\\
0 & \mbox{if $c>c_k$} \enspace .
\end{cases}
\]
The same counting argument here implies $c_k < 2^{k-1} \ln 2$,
while another simple algorithm yields $c_k = \Omega(2^k/k)$.
Again, no algorithm is known to improve this bound asymptotically,
leaving a multiplicative gap of order $\Theta(k)$ between the
upper and lower bound for this problem as well.
\medskip

In this paper, we use the {\em second moment method\/} to show
that random $k$-CNF formulas are satisfiable, and random
$k$-uniform hypergraphs are 2-colorable, for density up to
$2^{k-1} \ln 2-O(1)$. Thus, we determine the threshold for random
$k$-SAT within a factor of two and the threshold for Property B
within a small additive constant.

Recall that $F_k(n,rn)$ is \whp\ unsatisfiable if $r \ge  2^k \ln
2$. Our first main result is
\begin{theorem}
\label{thm:ksat} For all $k \geq 3$, $F_k(n,m=rn)$ is \whp\
satisfiable if $r\leq 2^{k-1}\ln 2 -2$.
\end{theorem}

Our second main result determines the Property B threshold within
an additive $1/2+o(1)$.
\begin{theorem}
\label{thm:hyp} {For all $k \geq 3$, 
$H_k(n,m=cn)$ is \whp\ non-2-colorable if}
\begin{equation}\label{eq:upper}
c > 2^{k-1} \ln 2 - \frac{\ln 2}{2}  \enspace .
\end{equation}
There exists a sequence $t_k \to 0$ such that for all $k \geq 3$, 
$H_k(n,m=cn)$ is \whp\ 2-colorable if
\begin{equation}\label{eq:lower}
c < 2^{k-1} \ln 2 - \frac{\ln 2}{2} - \frac{1+t_k}{2} \enspace .
\end{equation}
\end{theorem}

{The upper bound in~\eqref{eq:upper} corresponds to the density
for which the expected number of 2-colorings of $H_k(n,cn)$ is
$o(1)$.} Our main contribution is inequality~\eqref{eq:lower}
which we prove using the second moment method. In fact, our
approach {yields explicit bounds for the hypergraph 2-colorability
threshold for each value of $k$ (although ones that lack an
attractive closed form). We give the first few of these bounds in
Table~\ref{tab:val}.} We see that the gap between our upper and
lower bounds converges to its limiting value of $1/2$ rather
rapidly.
\begin{table}[h]\label{tab:val}
$$ \begin{array}{c|ccccccccc}
 k              & 3 & 4 & 5 & 6 & 7 & 8 & 9 & 10 & 11\\   \hline
\mbox{Upper Bound}
& 2.410 & 5.191 & 10.741 & 21.833 & 44.014 & 88.376 & 177.099 & 354.545 & 709.436\\
\mbox{Lower Bound}
& 1.5 & 4.083 & 9.973 & 21.190 & 43.432 & 87.827 & 176.570 & 354.027 & 708.925\\
\end{array}
$$ \caption{Bounds for the 2-colorability threshold of random
$k$-uniform hypergraphs.}
\end{table}


Unlike the algorithmic lower bounds for random $k$-SAT and
hypergraph 2-colorability, our arguments are non-constructive: we
establish that \whp\ solutions exist for certain densities but do
not offer any hint on how to find them. We believe that abandoning
the algorithmic approach for proving such lower bounds is natural
and, perhaps, necessary. At a minimum, the algorithmic approach is
limited to the small set of rather naive algorithms whose analysis
is tractable using current techniques. Perhaps more gravely, it
could be that {\em no\/} polynomial algorithm can overcome the
$\Theta(2^k/k)$ barrier. Determining whether this is true even for
certain limited classes of algorithms, \eg random walk
algorithms, is a very interesting open problem.

In addition, by not seeking out some specific truth assignment,
{as algorithms do}, the second moment method gives some first
glimpses of the ``geometry'' of the set of solutions. Deciphering
these first glimpses, getting clearer ones, and exploring
potential interactions between the geometry of the set of
solutions and computational hardness are great challenges lying
ahead.\medskip

We note that recently, and independently, Frieze and
Wormald~\cite{friezewormald} applied the second moment method to
random $k$-SAT in the case where $k$ is a moderately growing
function of $n$.  Specifically, they proved that when $k-\log_2 n
\rightarrow \infty$, $F_k(n,m)$ is \whp\ satisfiable if $m <
(1-\epsilon) m^*$ but \whp\ unsatisfiable if $m > (1+\epsilon)
m^*$, where $m^* =(2^k\ln 2-O(1)) \,n$ and
$\epsilon=\epsilon(n)>0$ is such that $\epsilon n \to \infty$.
Their result follows by a direct application of the second moment
method to the number of satisfying assignments of $F_k(n,m)$. As
we will see shortly, while this approach gives a very sharp bound
when $k-\log_2 n \rightarrow \infty$, it fails for any fixed $k$
and indeed for any $k = o(\log n)$. \medskip

We also note that since this work first
appeared~\cite{focs,usrandom}, the line of attack we {put forward
has had} several other successful applications. Specifically,
in~\cite{yuval}, the lower bound for the random $k$-SAT threshold
was improved to $2^k \ln 2 - O(k)$ by building on the insights
presented here. In~\cite{max}, the method was successfully
extended to random Max $k$-SAT~\cite{max}, while in~\cite{birk} it
was applied to random graph coloring. We discuss these subsequent
developments in the Conclusions.

\subsection{The second moment method and the role of symmetry}

The version of the second moment method we will use is given by
Lemma~\ref{lem:sec} below and follows from a direct application of
the Cauchy-Schwarz inequality (see \eg Remark 3.1 in~\cite{jlr}).
\begin{lemma}\label{lem:sec}
For any non-negative random variable $X$,
\begin{equation}\label{eq:second}
 \Pr[X > 0] \,\ge\, \frac{\ex[X]^2}{\ex[X^2]} \enspace .
\end{equation}
\end{lemma}

It is natural to try to apply Lemma~\ref{lem:sec} to random
$k$-SAT by letting $X$ be the number of satisfying truth
assignments of $F_k(n,m)$. Unfortunately, as we will see, this
``naive" application of the second moment method fails rather
dramatically: for all $k\geq 1$ and every $r > 0$, $\ex[X^2]
> (1+\beta)^n \,\ex[X]^2$ for some $\beta=\beta(k,r)>0$. As a result, the second moment
method only gives an exponentially small lower bound on the
probability of satisfiability.\medskip

The key step in overcoming this failure lies in realizing that we
are free to apply the second moment method to any random variable
$X$ such that $X>0$ implies that the formula is satisfiable. In
particular, we can let $X$ be the size of any {\em subset\/} of
the set of satisfying assignments. By choosing this subset
carefully, we can hope to significantly reduce the variance of $X$
relative to its expectation and use Lemma~\ref{lem:sec} to prove
that the subset is frequently non-empty. Indeed, we will establish
the satisfiability of random $k$-CNF by focusing on those
satisfying truth assignments {\em whose complement is also
satisfying}. In Section~\ref{sec:crapola} we will give some
intuition for why the number of such assignments has much smaller
variance than the number of all satisfying assignments. For now,
we observe that considering only such satisfying assignments is
equivalent to interpreting the random $k$-CNF formula $F_k(n,m)$
as an instance of Not-All-Equal (NAE) $k$-SAT, where a truth
assignment $\sigma$ is a solution if and only if under $\sigma$
every clause contains at least one satisfied literal {\em and\/}
at least one unsatisfied literal. In other words, our lower bound
for the $k$-SAT threshold in Theorem~\ref{thm:ksat} is, in fact, a
lower bound for the NAE $k$-SAT threshold.

Indeed, for both random NAE $k$-SAT and random hypergraph
2-colorability we will apply Lemma~\ref{lem:sec} naively, \ie by
letting $X$ be the number of solutions. This will give
Theorem~\ref{thm:hyp} and the values in Table~\ref{tab:val} for
hypergraph 2-colorability and, as we will see, exactly the same
bounds for random NAE $k$-SAT. (Indeed, the proof of
Theorem~\ref{thm:hyp} is a slight generalization of the proof for
random NAE $k$-SAT.) We will see that this success of the naive
second moment is due to the symmetry inherent in both problems,
\ie to the fact that the complement of a solution is also a
solution. Indeed, we feel that highlighting this role of symmetry
--- and showing how it can be exploited even in asymmetric problems
like $k$-SAT --- is our main conceptual contribution. Exploiting
these ideas in other Constraint Satisfaction problems that have a
permutation group acting on the variables' domain is an
interesting area for further research.

\subsection{Organization of the paper}

In Section~\ref{sec:back} we give some background on random
$k$-SAT and random hypergraph 2-colorability.  In
Section~\ref{sec:crapola} we explain why the second moment method
fails when applied to $k$-SAT directly, and give some intuition
for why counting only the NAE-satisfying assignments rectifies the
problem.  We also point out some connections to methods of
statistical physics. In Section~\ref{sec:ground} we lay the
groundwork for bounding the second moment for both NAE $k$-SAT and
hypergraph 2-colorability by dealing with some probabilistic
preliminaries, introducing a ``Laplace method" lemma for bounding
certain sums, and outlining our strategy. The actual bounding
occurs in Sections~\ref{sec:nae_bound} to~\ref{sec:gleas}.
{Specifically, in Sections~\ref{sec:nae_bound} and~\ref{sec:hyp}
we use the Laplace lemma to reduce the second moment calculations
for both random NAE $k$-SAT and random hypergraph 2-colorability
to the maximization of a certain function $g$ on the unit
interval, where $g$ is independent of $n$. We maximize $g$ in
Section~\ref{sec:gleas} and prove the Laplace lemma in
Section~\ref{sec:peak}.
We conclude in Section~\ref{sec:conc} by discussing some recent
extensions of this work and proposing several open questions.}

\section{Related Work}
\label{sec:back}

\subsection{Random $k$-SAT}
\label{sec:ksat_back}

The mathematical investigation of random $k$-SAT began with the
work of Franco and Paull~\cite{FrPa} who, among other results,
observed that $F_k(n,m=rn)$ is \whp\ unsatisfiable if $r \geq 2^k
\ln 2$. To see this, let $C_k = 2^k\binom{n}{k}$ be the number of
all possible $k$-clauses and let $S_k = (2^k-1)\binom{n}{k}$ be
the number of $k$-clauses consistent with a given truth
assignment. Since any fixed truth assignment is satisfying with
probability $\binom{S_k}{m}/\binom{C_k}{m}<(1-2^{-k})^m$, the
expected number of satisfying truth assignments of $F_k(n,m=rn)$
is at most $[2(1-2^{-k})^r]^n=o(1)$ for $r \geq 2^k \ln 2$.

Shortly afterwards, Chao and Franco~\cite{ChFrUC} complemented
this result by proving that for all $k \geq 3$, if $r < 2^k/k$ then the
following linear-time algorithm, called {\sc Unit Clause (uc)},
finds a satisfying truth assignment \wupp:  if there exist unit
clauses, pick one randomly and satisfy it; else pick a random
unset variable and give it a random value. Note that since {\sc
uc} succeeds only \wupp\ (rather than \whp) this does not imply a
lower bound for $r_k$.

The satisfiability threshold conjecture gained a great deal of
popularity in the early 1990s and has received an increasing
amount of attention since then. The polynomial-time solvable case
$k=2$ was settled early on: independently, Chv\'{a}tal and
Reed~\cite{mick}, Fernandez de la Vega~\cite{vega2sat}, and
Goerdt~\cite{GoTU} proved that $r_2=1$. Chv\'{a}tal and
Reed~\cite{mick}, in addition to proving $r_2=1$, gave the first
lower bound for $r_k$, strengthening the positive-probability
result of Chao and Franco~\cite{ChFrUC} by analyzing the following
refinement of {\sc uc}, called {\sc Short Clause (sc)}: if there
exist unit clauses, pick one randomly and satisfy it; else if
there exist binary clauses, pick one randomly and satisfy a random
literal in it;  else pick a random unset variable and give it a
random value. In~\cite{mick}, the authors showed that for all $k
\geq 3$, {\sc sc} finds a satisfying truth assignment \whp\ for $r
< (3/8) \,2^k/k$ and raised the question of whether this lower
bound for $r_k$ can be improved asymptotically.

A large fraction of the work on the satisfiability threshold
conjecture since then has been devoted to the first
computationally hard case, $k=3$, and a long series of
results~\cite{BFU,ChFrGUC,FrSu,pi,326,KKL_talk,dub_announ,Kap,JSV,KKKS,KMPS,EMFV,FrPa}
has narrowed the potential range of $r_3$. Currently this is
pinned between $3.52$ by Kaporis, Kirousis and
Lalas~\cite{KKL_talk} and Hajiaghayi and Sorkin~\cite{352} and
$4.506$ by Dubois and Boufkhad~\cite{dub_announ}.  Upper bounds
for $r_3$  come from probabilistic counting arguments, refining
the above calculation of the expected number of satisfying
assignments. Lower bounds, on the other hand, have come from
analyzing progressively more sophisticated algorithms.
Unfortunately, neither of these approaches helps narrow the
asymptotic gap between the upper and lower bounds for $r_k$. The
upper bounds only improve $r_k \leq 2^k \ln 2$ by a small additive
constant; the best algorithmic lower bound, due to Frieze and
Suen~\cite{FrSu}, is $r_k \geq a_k 2^k/k$ where $\lim_{k \to
\infty} a_k = 1.817...$

There are  two more results that stand out in the study of random
$k$-CNF formulas. In a breakthrough paper, Friedgut~\cite{frie}
proved the existence of a {\em non-uniform\/} satisfiability
threshold, \ie  of a sequence $r_k(n)$ around which the
probability of satisfiability goes from 1 to 0.
\begin{theorem}[\cite{frie}]
\label{thm:frie} For each $k \geq 2$, there exists a sequence
$r_k(n)$ such that for every $\epsilon>0$,
\[
\limninf \Pr[F_k(n,rn)\mbox{ is satisfiable}] =
\begin{cases}
1 & \mbox{if $r = (1-\epsilon)\, r_k(n)$}\\
0 & \mbox{if $r = (1+\epsilon)\, r_k(n)$} \enspace .
\end{cases}
\]
\end{theorem}

In~\cite{ChvSz}, Chv\'{a}tal and Szemer\'{e}di established a
seminal result in proof complexity, by extending the work of
Haken~\cite{Haken} and Urquhart~\cite{urq} to random formulas.
Specifically, they proved that for all $k \geq 3$, if $r \geq 2^k
\ln 2$ then \whp\ $\fo{k}{rn}$ is unsatisfiable but every
resolution proof of its unsatisfiability contains at least
$(1+\eps)^n$ clauses, for some $\eps = \eps(k,r)>0$.
In~\cite{abm}, Achlioptas, Beame and Molloy extended the main
result of~\cite{ChvSz} to random CNF formulas that also contain
\mbox{2-clauses,} as this is relevant for the behavior of
Davis-Putnam (DPLL) algorithms on random $k$-CNF. (DPLL algorithms
proceed by setting variables sequentially, according to some
heuristic, and backtracking whenever a contradiction is reached.)
By combining the results in the present paper with the results
in~\cite{abm}, it was recently  shown~\cite{largek} that a number
of DPLL algorithms require exponential time {\em significantly
below\/} the satisfiability threshold, \ie for provably
satisfiable random $k$-CNF formulas.

Finally, we note that if one chooses to live unencumbered by the
burden of mathematical proof, powerful non-rigorous techniques of
statistical physics, such as the ``replica method'', become
available.  Indeed, several claims based on the replica method
have been subsequently established rigorously, so it is frequently
(but definitely not always) correct. Using this technique,
Monasson and Zecchina~\cite{MZRK} predicted $r_k \simeq 2^k \ln
2$. Like most arguments based on the replica method, their
argument is mathematically sophisticated but far from
rigorous.  In particular, they argue that as $k$ grows large, the
so-called {\em annealed approximation} should apply.
This creates an analogy with the second moment method which we
discuss in Section~\ref{sec:replica}.

\subsection{Random hypergraph 2-colorability}

While Bernstein~\cite{bernstein} originally raised the
2-colorability question for certain classes of infinite set
families, Erd\H{o}s popularized the finite version of the
problem~\cite{beck_3_chrom_hyper,Erd63_normat,LLL,KLRH,KS,RadSri00,SSU}
and the hypergraph representation. Recall that a 2-uniform
hypergraph, \ie a graph, is 2-colorable if and only if it has no
odd cycle. In a random graph with $cn$ edges this occurs with
constant probability if and only if $c < 1/2$
(see~\cite{FlaKnuPit89} for more on the evolution of cycles in
random graphs).

For all $k \ge 3$, on the other hand, hypergraph 2-colorability is
NP-complete~\cite{Lov_hyper_complex} and determining the
2-colorability threshold $c_k$ for $k$-uniform hypergraphs
$H_k(n,cn)$ remains open. Analogously to random $k$-SAT, we will
take the liberty of writing $c_k \geq c^*$ if $H_k(n,cn)$ is
2-colorable \whp\ for all $c < c^*$, and $c_k \leq c^*$  if
$H_k(n,cn)$ is \whp\ non-2-colorable for all $c > c^*$.

Alon and Spencer~\cite{AloSpeun} were the first to give bounds on
the potential value of $c_k$. Specifically, they observed that,
analogously to random $k$-SAT, the expected number of 2-colorings
of $H_k(n,cn)$ is at most $[2(1-2^{1-k})^c]^n$ and concluded that
$H_k(n,cn)$ is \whp\ non-$k$-colorable if $c \geq 2^{k-1} \ln 2$.
More importantly, by employing the Lov\'asz local lemma, they
proved that $H_k(n,cn)$ is \whp\ 2-colorable if $c =O(2^k/k^2)$.
Regarding the upper bound, it is easy to see that, in fact,
$2(1-2^{1-k})^c<1$ if $c=2^{k-1}\ln 2 - (\ln 2)/2$ and this yields
the upper bound of Theorem~\ref{thm:hyp}. Moreover, the techniques
of~\cite{KKKS,dubo} can be used to improve this bound further to
$2^{k-1}\ln 2 - (\ln 2)/2 - 1/4+t_k$, where $t_k \to 0$.

The lower bound of~\cite{AloSpeun} was improved by Achlioptas,
Kim, Krivelevich and Tetali~\cite{hyp2col} motivated by the
analogies drawn in~\cite{AloSpeun} between hypergraph
2-colorability and earlier work~\cite{ChFrUC,mick} for random
$k$-SAT. Specifically, it was shown in~\cite{hyp2col} that a
simple, linear-time algorithm \whp\ finds a 2-coloring of
$H_k(n,cn)$ for $c =O(2^k/k)$, implying $c_k=\Omega(2^k/k)$. These
were the best bounds for $c_k$ prior to Theorem~\ref{thm:hyp} of
the present paper.

Finally, we note that Friedgut's result~\cite{frie} applies to
hypergraph 2-colorability as well, \ie
\begin{theorem}[\cite{frie}]
\label{thm:frie_hyp} For each $k \geq 3$, there exists a sequence
$c_k(n)$ such that for every $\epsilon>0$,
\[
\limninf \Pr[H_k(n,cn)\mbox{ is 2-colorable}] =
\begin{cases}
1 & \mbox{if $c = (1-\epsilon)\, c_k(n)$}\\
0 & \mbox{if $c = (1+\epsilon)\, c_k(n)$} \enspace .
\end{cases}
\]
\end{theorem}

%
%
%

\section{The second moment method: first look}
\label{sec:crapola}

In the rest of the paper it will be convenient to work with a
model of random formulas that differs slightly from $F_k(n,m)$.
Specifically, to generate a random $k$-CNF formula on $n$
variables with $m$ clauses we simply generate a string of $km$
independent random literals, each such literal drawn uniformly
from among all $2n$ possible ones. Note that this is equivalent to
selecting, with replacement, $m$ clauses from among all possible
$2^k n^k$ ordered $k$-clauses. This choice of distribution for
$k$-CNF formulas will simplify our calculations significantly. As
we will see in Section~\ref{sec:gene}, the derived results can be
easily transferred to all other standard models for random $k$-CNF
formulas.

\subsection{Random $k$-SAT}

For any formula $F$, given truth assignments
$\sigma_1,\sigma_2,\ldots \in \{0,1\}^n$ we will write
$\sigma_1,\sigma_2,\ldots \models F$ to denote that {\em all\/} of
$\sigma_1,\sigma_2,\ldots$  satisfy $F$. Let $X=X(F)$ denote the
number of satisfying assignments of a formula $F$. Then, for a
 $k$-CNF formula with random clauses $c_1,c_2,\ldots,c_m$ we have
\begin{equation}\label{eq:fmsat}
\ex[X] = \ex\left[\sum_{\sigma} \ones_{\sigma \models
F}\right] = \sum_{\sigma}\ex\left[\prod_{c_i}\ones_{\sigma \models
c_i} \right] = \sum_{\sigma}\prod_{c_i} \ex[\ones_{\sigma \models
c_i}] = 2^n (1-2^{-k})^{m} \enspace ,
\end{equation}
since clauses are drawn independently and the probability that
$\sigma$ satisfies the $i$th random clause, \ie
$\ex[\ones_{\sigma \models c_i}]$, is $1-2^{-k}$ for every
$\sigma$ and $i$. Similarly, for $\ex[X^2]$ we have
\begin{equation}\label{eq:smsat}
\ex[X^2] = \ex\left[\left(\sum_{\sigma} \ones_{\sigma \models
F}\right)^2\right] = \ex\left[\sum_{\sigma,\tau}
\ones_{\sigma,\tau \models F}\right] =
\sum_{\sigma,\tau}\ex\left[\prod_{c_i}\ones_{\sigma,\tau \models
c_i} \right] = \sum_{\sigma,\tau}\prod_{c_i}
\ex[\ones_{\sigma,\tau \models c_i}] \enspace .
\end{equation}
We claim that $\ex[\ones_{\sigma,\tau \models c_i}]$, \ie the
probability that a fixed pair of truth assignments $\sigma,\tau$
satisfy the $i$th random clause, depends only on the number of
variables $z$ to which $\sigma$ and $\tau$ assign the same value.
Specifically, if the overlap is  $z= \alpha n$, we claim that this
probability is
\begin{equation}
 f_S(\alpha) = 1 - 2^{1-k} + 2^{-k} \alpha^k \enspace .
\end{equation}
Our claim follows by inclusion-exclusion and observing that if
$c_i$ is not satisfied by $\sigma$, the only way for it to also
not be satisfied by $\tau$ is for all $k$ variables in $c_i$ to
lie in the overlap of $\sigma$ and $\tau$. Thus, $f_S$ quantifies
the correlation between the events that $\sigma$ and $\tau$ are
satisfying as a function of their overlap. In particular, observe
that truth assignments with overlap  $n/2$ are uncorrelated since
$f_S(1/2) = (1-2^{-k})^2 = \Pr[\sigma {\mbox{ is satisfying}}]^2$.

Since the number of ordered pairs of assignments with overlap $z$
is $ 2^n \,\binom{n}{z}$ we thus have
\begin{equation}
 \ex[X^2] = 2^n \sum_{z=0}^n  \binom{n}{z} \,f_S(z/n)^{m}
 \enspace .
\end{equation}
%
Writing $z=\alpha n$ and using  the approximation $\binom{n}{z} =
(\a^\a(1-\a)^{1-\a})^{-n} \times {\mathrm {poly}}(n)$ we see that
\[
\ex[X^2]  \geq   2^n \left(\max_{0 \leq \a \leq 1}
            \left[\frac{f_S(\a)^r}{\a^\a(1-\a)^{1-\a}}\right]
        \right)^n \times {\mathrm {poly}}(n)
 \equiv  \left(\max_{0 \leq \a \leq 1} \Lambda_S(\a)\right)^n
\times {\mathrm {poly}}(n) \enspace .
\]
At the same time observe that $\ex[X]^2 = \left(2^n
(1-2^{-k})^{rn}\right)^2 =
\left(4f_S(1/2)^r\right)^n=\Lambda_S(1/2)^n$. Therefore, if there
exists some $\a \in [0,1]$ such that
$\Lambda_S(\alpha)>\Lambda_S(1/2)$ then the second moment is
exponentially greater than the square of the expectation and we
only get an exponentially small lower bound for $\Pr[X>0]$. Put
differently, unless the dominant contribution to $\ex[X^2]$ comes
from ``uncorrelated" pairs of satisfying assignments, \ie pairs
with overlap $n/2$, the second moment method fails.

With these observations in mind, in Fig.~1 we plot
$\Lambda_S(\alpha)$ for $k=5$ and different values of $r$. We see
that, unfortunately, for all values of $r$ shown $\Lambda_S$ is
maximized at some $\a > 1/2$. If we look closely into the two
factors comprising $\Lambda_S$, the reason for the failure of the
second moment method becomes apparent: while the entropic factor
$\left(\a^{\a}(1-\a)^{1-\a}\right)^{-1}$ is symmetric around
$1/2$, the correlation function $f_S$ is strictly increasing in
$[0,1]$. Therefore, the derivative of $\Lambda_S$ is never 0 at
1/2, instead becoming 0 at some $\a > 1/2$ where the benefit of
positive correlation balances with the cost of decreased entropy.
{(Indeed, this is true for all $k = o(\log n)$ and constant
$r>0$.)}

\vspace*{1cm}

\begin{figure}[h]\label{fig_sat}
  \centerline{\psfig{figure=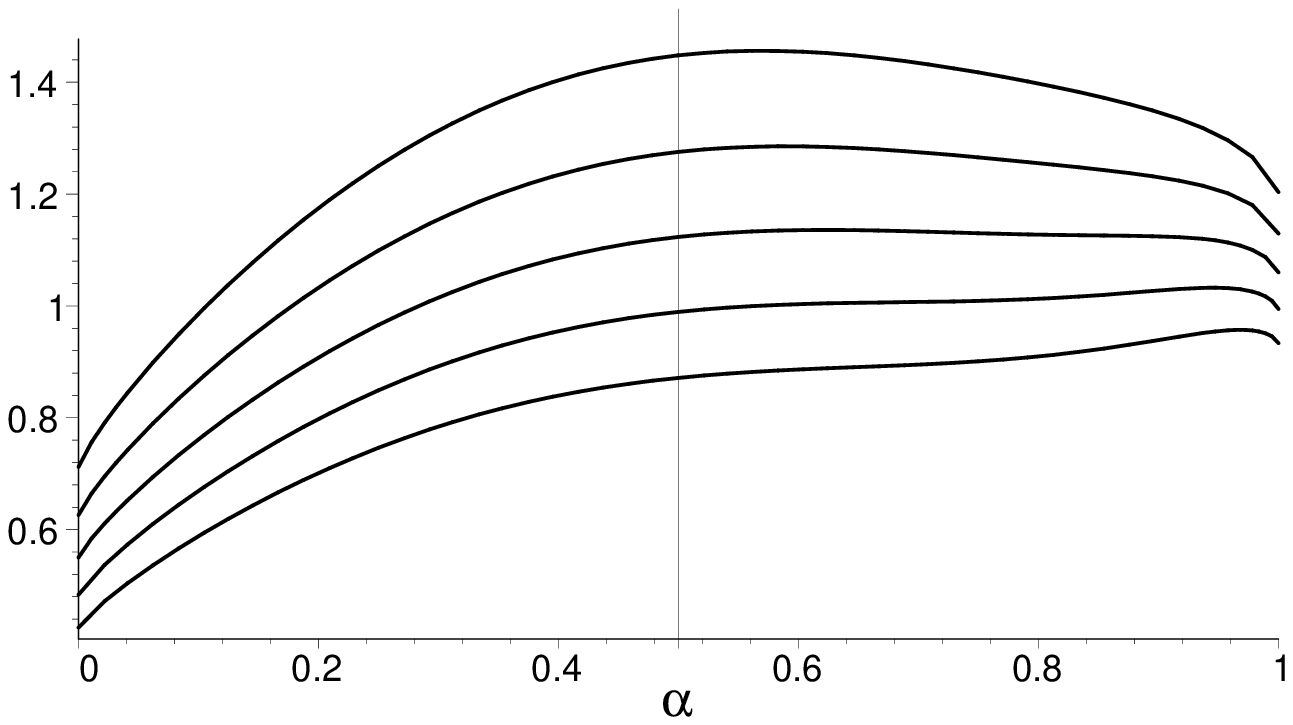,height=1.5in,width=3in}}
  \centerline{$k=5$, $r=16,18,20,22,24$ (top to bottom)}
\smallskip \caption{The contribution to $\ex[X^2]$ as a function of the
overlap for random $k$-SAT.}
\end{figure}

\subsection{Random NAE $k$-SAT}\label{july4}

Let us now repeat the above analysis but with $X=X(F)$ being the
number of NAE-satisfying truth assignments of a formula $F$.
Recall that $\sigma$ is a NAE-satisfying assignment iff under
$\sigma$ every clause has at least one satisfied literal {\em
and\/} at least one unsatisfied literal. Thus,  for a  $k$-CNF
formula with random clauses $c_1,c_2,\ldots,c_m$, proceeding as
in~\eqref{eq:fmsat}, we get
\begin{equation}
\ex[X] = 2^n (1-2^{1-k})^{m} \enspace ,
\end{equation}
since the probability that $\sigma$ NAE-satisfies the $i$th
random clause is $1-2^{1-k}$ for every $\sigma$ and $i$.

Regarding the second moment, proceeding exactly as
in~\eqref{eq:smsat}, we write $\ex[X^2]$ as a sum over the $4^n$
ordered pairs of assignments of the probability that both
assignments are NAE-satisfying. As for $k$-SAT, for any fixed pair
this probability depends only on the overlap. The only change is
that if $\sigma,\tau$ agree on $z=\alpha n$ variables then the
probability they both NAE-satisfy a random clause $c_i$ is
\begin{eqnarray}
\Pr[\mbox{$\sigma$ and $\tau$ NAE-satisfy $c_i$}] & = & 1-
2^{2-k}+2^{1-k}\left(\a^k+(1-\a)^k\right) \nonumber \\
& \equiv & f_N(\alpha)\enspace \label{eq:fn} \enspace .
\end{eqnarray}
Again, this claim follows from inclusion-exclusion and observing
that for $\sigma,\tau$ to both NAE-violate $c_i$, the variables of
$c_i$ must either all be in the overlap of $\sigma$ and $\tau$ or
all be in their non-overlap.

Applying Stirling's approximation for the factorial again and
observing that the sum defining $\ex[X^2]$ has only a polynomial
number of terms, we now get
\begin{equation}\label{eq:fulfill}
\ex[X^2]  \leq   2^n \left(\max_{0 \leq \a \leq 1}
            \left[\frac{f_N(\a)^r}{\a^\a(1-\a)^{1-\a}}\right]
        \right)^n \times {\mathrm {poly}}(n)
 \equiv  \left(\max_{0 \leq \a \leq 1} \Lambda_N(\a)\right)^n
\times {\mathrm {poly}}(n) \enspace .
\end{equation}

As before, it is easy to see that $\ex[X]^2 = \Lambda_N(1/2)^n$.
Therefore, if $\Lambda_N(1/2)>\Lambda_N(\alpha)$ for every $\a
\neq 1/2$ then~\eqref{eq:fulfill} implies that the ratio between
$\ex[X^2]$ and $\ex[X]^2$ is at most polynomial in $n$. Indeed,
with a more careful analysis of the interplay between the
summation and Stirling's approximation, we will later show that
whenever $\Lambda_N(1/2)$ is a global maximum, the ratio
$\ex[X^2]/\ex[X]^2$ is bounded by a constant, implying that
NAE-satisfiability holds \wupp\ So, all in all, again we hope that
the dominant contribution to $\ex[X^2]$ comes from pairs of
assignments with overlap $n/2$.

The crucial difference  is that now the correlation function $f_N$
is {\em symmetric\/} around $1/2$ and, hence, so is $\Lambda_N$.
As a result, the entropy-correlation product $\Lambda_N$ always
has a local extremum at $1/2$. Moreover, since the entropic term
is always maximized at $\alpha = 1/2$ and is independent of $r$,
for sufficiently small $r$ this extremum is a global maximum. With
these considerations in mind, in Fig.~2 we plot $\Lambda_N(\a)$
for $k=5$ and various values of $r$.

Let us start with the picture on the left, where $r$ increases
from 8 to 12 as we go from top to bottom. For $r=8, 9$ we see that
indeed $\Lambda_N$ has a global maximum at $1/2$ and the second
moment method succeeds. For the cases $r=11, 12$, on the other
hand, we see that $\Lambda_N(1/2)$ is actually a global minimum.
In fact, we see that $\Lambda_N(1/2)<1$, implying that $\ex[X]^2 =
\Lambda_N(1/2)^n =  o(1)$ and so \whp\ there are no NAE-satisfying
assignments for such $r$. It is worth noting that for $r=11$, even
though $X=0$ \whp, the second moment is exponentially large (since
$\Lambda_N>1$ near $0$ and $1$).

\begin{figure}\label{fig_nae}
\begin{minipage}{3in}
\begin{center}
  \centerline{\psfig{figure=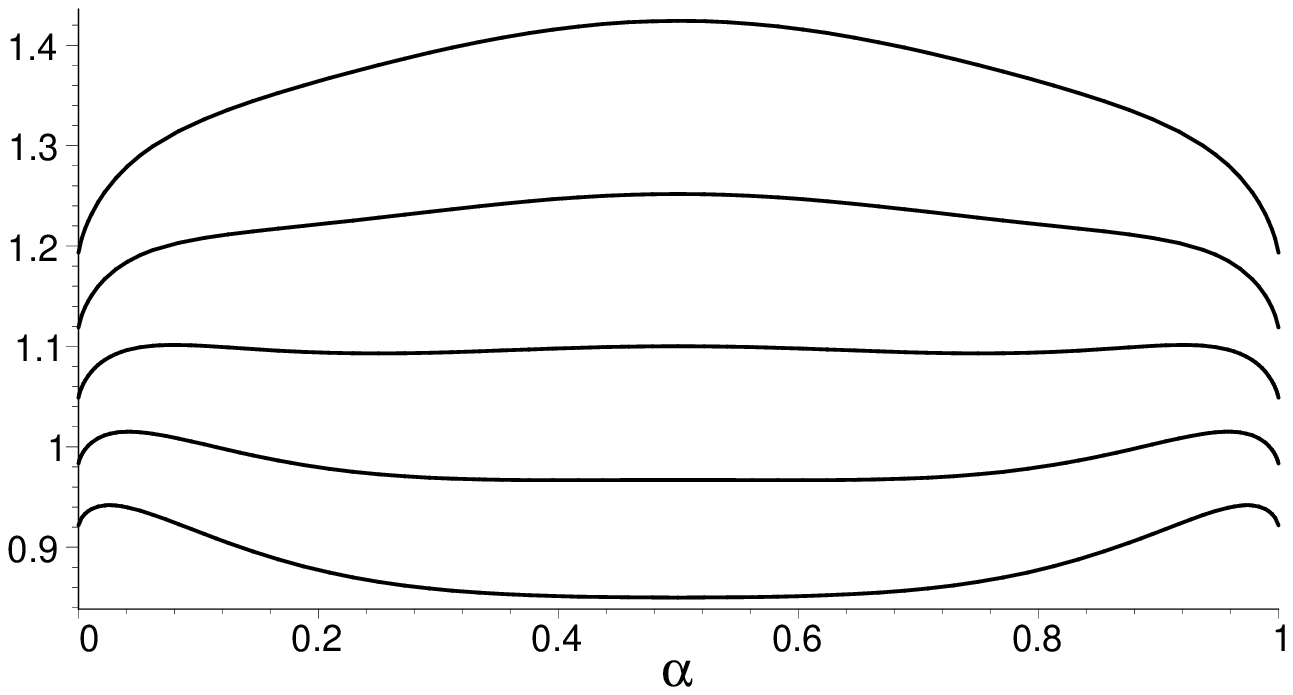,height=1.5in,width=2in}}
  \centerline{$k=5$, $r=8,9,10,11,12$ (top to bottom)}
\end{center}
\end{minipage}
\begin{minipage}{3in}
\begin{center}
  \centerline{\psfig{figure=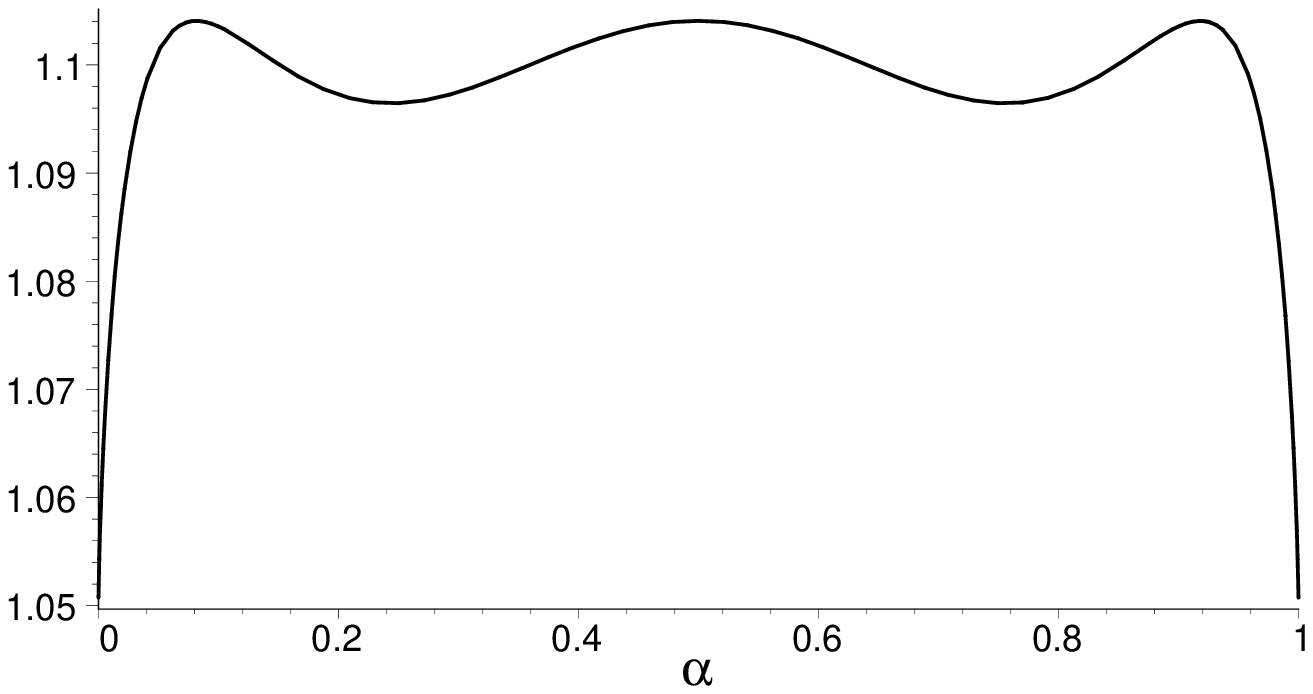,height=1.5in,width=2in}}
  \centerline{$k=5$, $r=9.973$}
\end{center}
\end{minipage}
\caption{The contribution to $\ex[X^2]$ as a function of the
overlap for random NAE $k$-SAT.}
\end{figure}
The most interesting case is $r=10$. Here $\Lambda(1/2)=1.0023...$
is a local maximum and greater than 1, but the two global maxima
occur at $\alpha = 0.08... $ and $\alpha=0.92...$ where the
function equals $1.0145$... As a result, again, the second moment
method only gives an exponentially small lower bound on $\Pr[X >
0]$. Note that this is in spite of the fact that $\ex[X]$ is now
exponentially large. Indeed, the largest value for which the
second moment succeeds for $k=5$ is $r = 9.973...$ when the two
side peaks reach the same height as the peak at $1/2$ (see the
plot on the right in  Fig.\ 2).

So, the situation can be summarized as follows. By requiring that
we only count NAE-satisfying truth assignments we make it,
roughly, twice as hard to satisfy each clause. This manifests
itself in the additional factor of 2 in the middle term of $f_N$
compared to $f_S$. On the other hand, now, the third term of $f$,
capturing ``joint" behavior, is symmetric around $1/2$, making
$\Lambda$ itself symmetric around $1/2$.  {This enables the second
moment method which, indeed, only breaks down when the density
gets within an additive constant of the upper bound for the NAE
$k$-SAT threshold}.


\subsection{How symmetry reduces variance}
\label{sec:boost}

Given a truth assignment $\sigma$ and an arbitrary CNF formula
$F$, let $Q=Q(\sigma,F)$ denote the total number of literal {\em
occurrences\/} in $F$ satisfied by $\sigma$. So, for example, $Q$
is maximized by those truth assignments that assign every variable
its ``majority'' value. With this definition at hand, a potential
explanation of how symmetry reduces the variance is suggested by
considering the following trivial refinement of our generative
model: 
first i) draw $km$ i.i.d.\ uniformly random literals just as
before and then ii) partition the drawn literals randomly into
$k$-clauses (rather than assuming that the first $k$ literals form
the first clause, the next $k$ the second, etc.).

In particular, imagine that we have just finished performing the
first generative step above and we are about to perform the
second. Observe that at this point the value of $Q$ has already
been determined for every $\sigma \in \{0,1\}^n$. Moreover, for
each fixed $\sigma$ the conditional probability of yielding a
satisfying assignment corresponds to a balls-in-bins experiment:
distribute $Q(\sigma)$ balls in $m$ bins, each with capacity $k$,
so that every bin receives at least one ball. It is clear that
those truth assignments for which $Q$ is large at the end of the
first step have a big advantage in the second.

To get an idea of what $Q$ typically looks like on $\{0,1\}^n$ we
begin by observing that the number of occurrences of a fixed
literal $\ell$, $B_{\ell}$,  is distributed as
${\mathrm{Bin}}(km,1/(2n))$. Thus, $\ex[B_{\ell}] = O(1)$ and,
moreover,  the random variables $B_{\ell}$ are very weakly
correlated. As a result, at the end of the first step, $Q$ is
typically a very smooth function on $\{0,1\}^n$, attaining a
maximum value at the subcube of majority vote assignments  and
gradually decreasing away from them. Thus, at the end of the first
step the ``more promising" truth assignments are correlated: in
satisfying many literal occurrences (thus increasing their odds
for the second step), they tend to overlap with each other (and
the majority assignment) at more than half the variables.

In contrast, if we focus on NAE-satisfying assignments, at the end
of the first step the most promising assignments $\sigma$ are
those for which $Q(\sigma)$ is very close to its average value
$km/2$. So, when the problem is symmetric, the typical case
becomes the most favorable case and the clustering around truth
assignments that satisfy many literal occurrences disappears.

If indeed ``populism'', \ie the tendency of each variable to
assume its majority value in the formula, is the main source of
correlations in random $k$-SAT, then the second moment method is a
good candidate for $k$-CNF models which do not encourage  this
tendency\footnote{We describe recent developments on this point
in the Conclusions.}.
For example, one such model is
{\em regular} random $k$-SAT, in which every literal occurs
exactly the same number of times. Such formulas can be analyzed
using a model analogous to the configuration model of random
graphs, \ie by taking precisely $d$ copies of each literal and
partitioning the resulting $2dn$ copies into clauses randomly
(exactly as in the second step of our two-step model for random
$k$-SAT).

\subsection{Geometry and connections to statistical physics} \label{sec:replica}

\comment{A quantity of interest in statistical physics is the {\em
overlap distribution}, namely the probability $P(\alpha)$ that a
random pair of satisfying assignments of a random formula has
overlap $\alpha$.  Our calculation of $\ex[X^2]$ amounts to
calculating $P(\alpha)$ in the so-called ``annealed
approximation'' where we average over random formulas first, in
which case $P(\alpha)$ is proportional to $g(\alpha)^n$.  While
this approximation gives a larger weight to formulas with more
satisfying assignments, Monasson and Zecchina~\cite{MZRK}
conjectured that it becomes tight in the limit of large $k$.  We
have shown that for small enough $r$, $g(\alpha)$ has a unique
maximum, and so $P(\alpha)$ is tightly peaked in the limit $n \to
\infty$.  This is consistent with the assumptions behind the
non-rigorous ``replica trick'' of statistical physics.  Moreover,
the fact that this peak occurs at $\alpha = 1/2$ shows that, for
these values of $r$, the satisfying assignments are scattered
throughout the hypercube {\em as if they were independent.}

Statistical physicists have developed a number of methods for
investigating phase transitions which, while non-rigorous, are
often in spectacular agreement with numerical and experimental
results. One of these is the replica method. The term ``replica''
comes from the fact that when $q$ is an integer one can compute
$\ex[X^q]$ by considering the interactions between $q$-tuples of
the objects counted by $X$. In physical terms, one creates $q$
replicas of the system (in our case of the formula) and considers
the joint probability distribution for $q$-tuples of states
(assignments) from different replicas. So, in these terms, our
calculation of the second moment involves looking at the
interactions between two replicas. At a high level, the replica
method amounts to computing $\ex[\ln X]$ by calculating $\ex[X^q]$
for all {\em integer\/} $q$ and then plugging in the resulting
formula to the expression $\ex[\ln X] = \lim_{q\to 0}{(\ex[X^q] -
1)}/q$. The fundamental leap of faith, of course, lies in allowing
the analytic continuation $q \to 0$ from integer values of $q$.
Even to get this far, however, one has to deal with the often
daunting task of computing $\ex[X^q]$ for all integer $q$.

Let us try to do this when $X$ counts satisfying assignments.
Similarly to our treatment for $\ex[X^2]$,  to calculate
$\ex[X^q]$ we can partition the $(2^n)^q$ terms in the sum
according to the ``overlap" between the $q$ truth assignments.
Specifically, this overlap is characterized by $2^q$ numbers, each
such number counting the number of variables to which the
corresponding 0/1 pattern of values occurs in the $q$ truth
assignments. By applying again Stirling's approximation we are
left to maximize a function of $2^q$ variables. Of course, since
these numbers must sum to $n$, this becomes a $(2^q -
1)$-dimensional maximization; moreover, in problems like $k$-SAT
or NAE $k$-SAT where every truth assignment has the same a priori
probability of being a solution, we can take one of the $q$
assignments to be all 0s, reducing this dimensionality slightly to
$2^{q-1}-1$ dimensions. Maximizing families of functions over an
exponential number of dimensions, however, is no small task.

By taking another leap of faith, one can reduce the dimensionality
of this maximization to $q$ by assuming {\em replica symmetry\/},
\ie that the global maximum is symmetric under permutations of the
replicas. In other words, in our problem this is the assumption
that the global maximum occurs when all overlap variables having
the same number of 1s take the same value. This reduces the
dimensionality from $2^q$ to $q+1$. Even when this assumption is
wrong (and it often is), it can lead to good approximations.  In
particular, Monasson and Zecchina~\cite{MZRK} used replica
symmetry to predict $r_k \simeq 2^k \ln 2$.

A standard indicator of the plausibility of replica symmetry in
given by the distribution of overlaps between randomly chosen
``ground states'', in our case satisfying assignments. If replica
symmetry holds, this overlap distribution is tightly peaked around
its mean; if not, it typically gains multiple peaks or has support
on an open interval.  This ``replica symmetry breaking'' indicates
that rather than being smoothly distributed, ground states are
organized into clumps, with large overlaps within clumps and
smaller overlaps between them.

Intriguingly, the second-moment method is a calculation of the
overlap distribution in the {\em annealed approximation}, a common
approximation in physics. That is, when we compute this
distribution en route to computing $\ex[X^2]$ we are considering
all pairs of solutions and over all formulas (or, equivalently,
over a uniformly random formula). As a result, formulas with many
solutions have a stronger effect on the distribution that what
they should. While this approximation gives formulas with more
satisfying assignments a heavier influence, potentially throwing
averages off their correct values, on physical grounds Monasson
and Zecchina~\cite{MZRK} conjectured that the annealed
approximation should apply to $k$-SAT in the limit of large $k$.

We saw in the previous section that the entropy-correlation
product for NAE $k$-SAT is peaked at $\alpha = 1/2$ almost all the
way to the threshold.  When we take $n$th powers of this product
the contribution of all other terms vanishes, so the overlap
distribution is sharply concentrated around $\alpha = 1/2$.
Physically, we can thus say the following: at least in the
annealed approximation, the overlap distributions for NAE $k$-SAT
and hypergraph 2-colorability behave {\em as if the solutions to
these problems were scattered independently throughout the
hypercube.} As we saw above, this effective independence is what
reduces the variance and allows the second moment method to work.
}

A key quantity in statistical physics is the {\em overlap
distribution} between configurations of minimum energy, known as
ground states.  When a constraint satisfaction problem is satisfiable,
ground states correspond to solutions, such as satisfying assignments,
2-colorings, and so on.
In the case of random \mbox{$k$-SAT}, the overlap distribution is the probability
$P(\alpha)$ that a random pair of satisfying assignments
have overlap $\alpha n$.

This overlap distribution gives an intriguing interpretation of our results.
In writing $\ex[X^2]$ as a sum of contributions from pairs of assignments
with different overlaps, we have in fact calculated the
average of $P(\alpha)$ over all formulas, weighted by the number
of pairs of satisfying assignments of each one.
Physicists call this weighted average the ``annealed
approximation'' of $P(\alpha)$, and denote it $P_\ann(\alpha)$.
It is worth pointing out that, while the annealed approximation
clearly overemphasizes formulas with more satisfying assignments,
Monasson and Zecchina conjectured in~\cite{MZRK}, based on the
replica method, that it becomes asymptotically tight as $k \to
\infty$.

On a more rigorous footing, it is easy to see that in our case
$P_\ann(\alpha)$ is proportional to $\Lambda(\alpha)^n$.
Therefore, whenever $\Lambda$ is peaked at $1/2$, $P_\ann(\alpha)$
is tightly peaked around $1/2$, since $\Lambda(\alpha)^n$ vanishes
for all other values of $\alpha$ as $n \to \infty$. This is
precisely what we prove occurs in random \mbox{NAE $k$-SAT} for
densities up to $2^{k-1} \ln 2 - O(1)$.  In other words, for
densities almost all the way to the random NAE $k$-SAT threshold,
in the annealed approximation, the NAE-satisfying assignments are
scattered throughout the hypercube {\em as if they were
independent.}\medskip

{Note that even if $P(\alpha)$ is concentrated around 1/2 (rather
than just $P_\ann(\alpha)$) this still allows for a typical
geometry where there are exponentially many, exponentially large
clusters, each centered at a random assignment. Indeed, this is
precisely the picture suggested by some very recent,
ground-breaking work of Mezard, Parisi, and
Zecchina~\cite{MPZ,MZ}, based on non-rigorous techniques of
statistical physics. If this is indeed the true picture,
establishing it rigorously would require considerations much more
refined than the second moment of the number of solutions. More
generally, getting a better understanding of the typical geometry
and its potential implications for algorithms appears to us a very
challenging and very important open problem.}


\section{Groundwork}\label{sec:ground}

\subsection{Generative Models}\label{sec:gene}

Given a set $V$ of $n$ Boolean variables, let $C_k=C_k(V)$ denote
the set of all proper {\em $k$-clauses\/} on $V$, \ie the set of
all $2^k \binom{n}{k}$ disjunctions of $k$ literals involving
distinct variables. Similarly, given a set $V$ of $n$ vertices,
let $E_k=E_k(V)$ be the set of all $\binom{n}{k}$ $k$-subsets of
$V$. As we saw, a random $k$-CNF formula $F_k(n,m)$ is formed by
selecting uniformly a random $m$-subset of $C_k$, while a random
$k$-uniform hypergraph $H_k(n,m)$ is formed by selecting uniformly
a random $m$-subset of $E_k$.

While $F_k(n,m)$ and $H_k(n,m)$ are perhaps the most natural
models for generating random $k$-CNF formulas and random
$k$-uniform hypergraphs, respectively, there are a number of
slight variations of each model. Those are largely motivated by
amenability to certain calculations. To simplify the discussion we
focus on models for random formulas in the rest of this
subsection. All our comments transfer readily to models for random
hypergraphs.

For example, it is fairly common to consider the clauses as
ordered $k$-tuples (rather than as $k$-sets) and/or to allow
replacement in sampling the set $C_k$. Clearly, for properties
such as satisfiability the issue of ordering is irrelevant.
Moreover, as long as $m=O(n)$, essentially the same is true for
the issue of replacement. To see that, observe that \whp\ the
number of repeated clauses is $q=o(n)$ and the subset of $m-q$
distinct clauses is uniformly random. Thus, if a monotone
decreasing property (such as satisfiability) holds with
probability $p$ for a given $m=r^*n$ when replacement is allowed,
it holds with probability $p-o(1)$ for all $r<r^*$ when replacement
is not allowed.

The issue of selecting the literals of each clause with
replacement (which might result in some ``improper" clauses) is
completely analogous. That is, the probability that a variable
appears more than once in a given clause is at most $k^2/n=O(1/n)$
and hence \whp\ there are $o(n)$ improper clauses. Finally, we
note that by standard techniques our results also transfer to the
$F_k(n,p)$ model where every clause appears independently of all
others with probability $p$. For that it suffices to set $p$ such
that $2^k \binom{n}{k} p = r^*n - o(n)$.

\subsection{Strategy and Tools}
Our plan is to consider random $k$-CNF formulas formed by
generating $km$ i.i.d.\ random literals, where $m=rn$, and proving
that if $X=X(F)$ is the number of NAE-satisfying assignments then:
\begin{lemma}
\label{lem:naepos} For all $\eps > 0$, $k \geq k_0(\eps)$ and $r <
2^{k-1} \ln 2 - (1+\ln 2)/2 - \eps$, there exists some constant $C
= C(k,r) > 0$ such that
\[
{\ex[X^2]} < C \times {\ex[X]^2} \enspace .
\]
\end{lemma}

By Lemma~\ref{lem:sec} and our discussion in
Section~\ref{sec:gene}, this implies that $F_k(n,rn-o(n))$ is
NAE-satisfiable \wupp\ Since a NAE-satisfiable formula is also
satisfiable, we have established that $F_k(n,rn)$ is satisfiable
\wupp\ for all $r$ as in Lemma~\ref{lem:naepos}. To boost this to
a high probability result, thus establishing
Theorem~\ref{thm:ksat}, we employ the following immediate
corollary of Theorem~\ref{thm:frie}.
\begin{corollary}\label{cor:boost_nae}
If $F_k(n,r^*n)$ is satisfiable \wupp\ then $F_k(n,rn)$ is
satisfiable \whp\ for all $r < r^*$.
\end{corollary}

Friedgut's arguments~\cite{frie} apply equally well to NAE
$k$-SAT, implying that $F_k(n,rn)$ is \whp\ NAE-satisfiable for
$r$ as in Lemma~\ref{lem:naepos}. Thus, Lemma~\ref{lem:naepos}
readily yields~\eqref{eq:lower_nae} below,
while~\eqref{eq:upper_nae} comes from noting that the expected
number of NAE-satisfying assignments is $[2(1-2^{1-k})^r]^n$.
{(Similarly to hypergraphs, the techniques of~\cite{KKKS,dubo} can
be used to improve the bound in~\eqref{eq:upper_nae} to
$2^{k-1}\ln 2 - (\ln 2)/2 - 1/4+t_k$, where $t_k \to 0$.) Indeed,
we will see that the proof of Theorem~\ref{thm:naeksat} will yield
Theorem~\ref{thm:hyp} for random hypergraphs with little
additional effort. }

\begin{theorem}\label{thm:naeksat}
{For all $k \geq 3$, 
$F_k(n,m=rn)$ is \whp\ non-NAE-satisfiable if}
\begin{equation}\label{eq:upper_nae}
r > 2^{k-1} \ln 2 - \frac{\ln 2}{2} \enspace .
\end{equation}
There exists a sequence $t_k \to 0$ such that for all $k \geq 3$,
$F_k(n,m=rn)$ is \whp\ NAE-satisfiable if
\begin{equation}\label{eq:lower_nae} r < 2^{k-1} \ln 2 -
\frac{\ln 2}{2} - \frac{1+t_k}{2} \enspace
\end{equation}
\end{theorem}
As we saw in Section~\ref{july4}, the second moment of the number
of NAE-satisfying assignments is
\[
2^n \sum_{z=0}^n  \binom{n}{z} \,f_N(z/n)^{rn}
 \enspace .
\]
A slightly more complicated sum will occur when we bound the
second moment of the number of 2-colorings. To bound both sums we
will use the following lemma which we prove in
Section~\ref{sec:peak}.
\begin{lemma}
\label{lem:peak} {\bf [Laplace lemma]} Let $\phi$ be a positive,
twice-differentiable function on $[0,1]$ and let $q \geq 1$ be a
fixed integer. Let $t=n/q$ and let
\[
S_n =  \sum_{z=0}^{t} \binom{t}{z}^{\!q} \,\phi(z/t)^n \enspace .
\]
Letting $0^0 \equiv 1$, define $g$ on $[0,1]$ as
\[
g(\alpha) = \frac{\phi(\alpha)}
                 {\alpha^\alpha \,(1-\alpha)^{1-\alpha}} \enspace .
\]
 If there exists $\amax \in (0,1)$ such that
$g(\amax) \equiv \gmax > g(\alpha)$ for all $\alpha \neq \amax$,
and $g''(\amax) < 0$, then there exists a constant
$C=C(q,\amax,\gmax,g''(\amax))>0$ such that
\[
S_n < C \,{n^{-(q-1)/2} \,\gmax^n} \enspace .
\]
\end{lemma}

\section{Bounding the Second Moment for NAE
$k$-SAT}\label{sec:nae_bound}

Recall that if $X$ is the number of NAE-assignments, then
\[
\ex[X] = 2^n (1-2^{1-k})^{rn}
\]
and
\begin{equation}\label{lau}
\ex[X^2] = 2^n \sum_{z=0}^n  \binom{n}{z} \,f_N(z/n)^{rn} \enspace
,
\end{equation}
where
$$
f_N(\alpha) = 1- 2^{2-k}+2^{1-k}\left(\a^k+(1-\a)^k\right)
\enspace .
$$
To bound the sum in~\eqref{lau} we  apply Lemma~\ref{lem:peak}
with $q=1$ and $\phi(\alpha) = f_N(\alpha)^r$. Thus, 
$g = g_r$ where
\begin{equation}\label{lotr}
g_r(\alpha) =
\frac{f_N(\alpha)^r}{\alpha^{\alpha}(1-\alpha)^{1-\alpha}}
\enspace.
\end{equation}
To show that Lemma~\ref{lem:peak} applies, we will prove in
Section~\ref{sec:gleas} that
\begin{lemma}\label{gleas}
For every $\eps > 0$, there exists $k_0=k_0(\eps)$ such that for
all $k \geq k_0$, if
\[ r < 2^{k-1} \ln 2 - \frac{\ln 2}{2} - \frac{1}{2} - \eps \]
then $g_r(\alpha) < g_r(1/2)$ for all $\alpha \neq 1/2$, and
$g_r''(1/2) < 0$.
\end{lemma}
Therefore, for all $r$, $k$, and $\eps$ as in Lemma~\ref{gleas},
there exists a constant $C = C(k,r) > 0$ such that
\[
{\ex[X^2]} < C \times {2^n g_r(1/2)^n} \enspace .
\]
Since $\ex[X]^2 =  2^n g_r(1/2)^n$ we get that for all $r, k,
\eps$ as in Lemma~\ref{gleas}
\[
{\ex[X^2]}< C \times {\ex[X]^2} \enspace .
\]

\section{Bounding the Second Moment for  Hypergraph 2-colorability}
\label{sec:hyp}

Just as for NAE $k$-SAT, it will be easier to work with the model
in which generating a random hypergraph corresponds to generating
$km$ random vertices, each such vertex chosen uniformly at random
with replacement, and letting the first $k$ vertices form the
first hyperedge etc.

In~\cite{usrandom} we proved \eqref{eq:lower} of
Theorem~\ref{thm:hyp} by letting $X$ be the set of all 2-colorings
and using a convexity argument to show that $\ex[X^2]$ is
dominated by the contribution of {\em balanced} colorings, \ie
colorings with an equal number of black and white vertices.  Here
we follow a simpler approach suggested by David Karger; namely, we
{\em define\/} $X$ to be the number of balanced 2-colorings. We
emphasize that, while technically convenient, the restriction to
balanced 2-colorings is not essential for the second moment method
to succeed on hypergraph 2-colorability, \ie one has $\ex[X^2] =
O(\ex[X]^2)$ even if $X$ is the number of all 2-colorings.

Of course, in order for balanced colorings to exist $n$ must be
even and we will assume that in our calculations below. To get
Theorem~\ref{thm:hyp} for all sufficiently large $n$, we observe
that if for a given $c^*$, $H_k(2n,m=2c^*n)$ is 2-colorable \whp\
then for all $c<c^*$, $H_k(n,cn)$ is 2-colorable \whp\ since
deleting a random vertex of $H_k(2n,2c^*n)$ \whp\ removes $o(n)$
edges. With this in mind, in the following we let $X$ be the
number of balanced 2-colorings and assume that $n$ is even.

Since the vertices in each hyperedge are chosen uniformly with
replacement, then the probability that a random hyperedge is
bichromatic in a fixed balanced partition is $1-2^{1-k}$. Since
there are $\binom{n}{n/2}$ such partitions and the $m$ hyperedges
are drawn independently, we have
\begin{equation}
\label{eq:hypfirst} \ex[X] = \binom{n}{n/2}
\,\left(1-2^{1-k}\right)^{m} \enspace .
\end{equation}

To calculate the second moment, as we did for [NAE] $k$-SAT, we
write $\ex[X^2]$ as a sum over all pairs of balanced partitions.
In order to estimate this sum we first observe that if two
balanced partitions $\sigma$ and $\tau$ have exactly $z$ black
vertices in common, then they must also have exactly $z$ white
vertices in common. Thus $\sigma$ and $\tau$ define four groups of
vertices: $z$ that are black in both, $z$ that are white in both,
$n/2-z$ that are black in $\sigma$ and white in $\tau$, and
$n/2-z$ that are white in $\sigma$ and black in $\tau$. Clearly, a
random hyperedge is monochromatic in both $\sigma$ and $\tau$ iff
all its vertices fall into the same group. Since the vertices of
each hyperedge are chosen uniformly with replacement, this
probability is
\[
2 \left( \frac{z}{n} \right)^k + 2 \left( \frac{n/2-z}{n} \right)^k
= 2^{1-k} \left[ \left( \frac{2z}{n} \right)^{\!k} + \left( 1 - \frac{2z}{n} \right)^{\!k} \right]
\enspace .
\]
Thus, by inclusion-exclusion, the probability that a random
hyperedge is bichromatic in both $\sigma$ and $\tau$ is
\[
1 - 2^{2-k} + 2^{1-k} \left[ \left( \frac{2z}{n} \right)^{\!k} +
\left( 1 - \frac{2z}{n} \right)^{\!k} \right] = f_N(2z/n)
\]
where $f_N(\alpha)=1 - 2^{2-k} + 2^{1-k} (\alpha^k +
(1-\alpha)^k)$ is the function we defined for NAE $k$-SAT
in~\eqref{eq:fn}.

Moreover, observe that the number of pairs of partitions with such
overlap is
\[
\binom{n}{z, \,z, \,n/2-z, \,n/2-z} = \binom{n}{n/2}
\binom{n/2}{z}^{\!2} \enspace .
\]
Since hyperedges are drawn independently and with replacement, by
summing over $z$ we thus  get
\[
\ex[X^2]  =  \binom{n}{n/2} \,\sum_{z=0}^{n/2} \binom{n/2}{z}^{\!2}
   f_N(2z/n)^{cn} \enspace .
\]
To bound this sum we apply Lemma~\ref{lem:peak} with $q=2$ and
$\phi(\alpha) = f_N(\alpha)^c$.  Felicitously, we find ourselves
maximizing a function $g_c$ which, if we replace $c$ with $r$,
is exactly the same function $g_r$ we defined
in~\eqref{lotr} for NAE $k$-SAT.  Thus, setting $c=r$ where $k,r$ and $\eps$
are as in Lemma~\ref{gleas}, $g_c$ is maximized at $\alpha = 1/2$ with
$g''(1/2) < 0$, and Lemma~\ref{lem:peak} implies that there exists
a constant $C=C(r,k) > 0$ such that
\[
{\ex[X^2]} < C \, {n^{-1/2} \, \binom{n}{n/2} \,g_c(1/2)^n}
\enspace .
\]

We now bound $\ex[X]$ from below using Stirling's
approximation~\eqref{eq:stirling} and get
\[
\frac{\ex[X^2]}{\ex[X]^2} < C \times \frac{n^{-1/2}
\binom{n}{n/2} \,g_c(1/2)^n}{{\binom{n}{n/2}}^2
    \,(1-2^{1-k})^{2rn}}
= C \times \frac{n^{-1/2}\,2^n}{\binom{n}{n/2}} \to C \times
\sqrt{\frac{\pi}{2}} \enspace .
\]
To complete the proof, analogously to [NAE] $k$-SAT, we use the
following ``boosting" corollary of Theorem~\ref{thm:frie_hyp}.
\begin{corollary}\label{cor:boost_hyp}
If $H_k(n,c^*n)$ is 2-colorable \wupp\ then $H_k(n,cn)$ is
2-colorable \whp\ for all $c < c^*$.
\end{corollary}

\section{Proof of Lemma~\ref{gleas}}
\label{sec:gleas}

We need to prove $g''_r(1/2) < 0$ and $g_r(\alpha) < g_r(1/2)$ for
all $\alpha \ne 1/2$.  As $g_r$ is symmetric around $1/2$, we can
restrict to $\alpha \in (1/2,1]$. We divide $(1/2,1]$ into two
parts and handle them with two separate lemmata. The first lemma
deals with $\alpha \in (1/2,0.9]$ and also establishes that
$g''_r(1/2) < 0$.
\begin{lemma}
\label{lem:nearhalf} Let $\alpha \in (1/2,0.9]$. For all $k \ge
74$, if $r \leq 2^{k-1} \ln 2$ then $g_r(\alpha) < g_r(1/2)$ and
$g_r''(1/2) < 0$.
\end{lemma}
The second lemma deals with $\alpha \in (0.9,1]$.
\begin{lemma}
\label{lem:far} Let $\alpha \in (0.9,1]$.  For every $\eps > 0$
and all $k \geq k_0(\eps)$, if $r \leq 2^{k-1} \ln 2 - \frac{\ln
2}{2} - \frac{1}{2} - \eps$ then $g_r(\alpha) < g_r(1/2)$.
\end{lemma}
Combining Lemmata~\ref{lem:nearhalf} and \ref{lem:far} we see that
for every $\eps > 0$ and $k \ge k_0=k_0(\eps)$,  if
\[ r \leq 2^{k-1} \ln 2 - \frac{\ln 2}{2} - \frac{1}{2} - \eps \]
then $g_r(\alpha) < g_r(1/2)$ for all $\alpha \ne 1/2$ and
$g_r''(1/2)<0$, establishing Lemma~\ref{gleas}.

We prove Lemmata \ref{lem:nearhalf} and~\ref{lem:far} below. The
reader should keep in mind that we have made no attempt to
optimize the value of $k_0$ in Lemma~\ref{lem:far}, aiming instead
for proof simplicity. For the lower bounds presented in
Table~\ref{tab:val} we computed numerically, for each $k$, the
largest value of $r$ for which the conclusions of
Lemma~~\ref{gleas} hold. In each case, the condition $g''(1/2)<0$
was satisfied with room to spare, while establishing
$g(1/2)>g(\alpha)$ for all $\alpha \neq 1/2$ was greatly
simplified by the fact that $g$ always has no more than three
local extrema in $[0,1]$.

\medskip

\noindent {\bf Proof of Lemma~\ref{lem:nearhalf}.} We will first
prove that for $k \geq 74$, $g_r$ is strictly decreasing in
$\alpha = (1/2,0.9]$, thus establishing $g_r(\alpha)<g_r(1/2)$.
Since $g_r$ is positive, to do this it suffices to prove that
$(\ln g_r)' = g'_r/g_r < 0$ in this interval. In fact, since
$g'_r(\alpha) = (\ln g_r)' = 0$ at $\alpha = 1/2$, it will suffice
to prove that for  $\alpha \in [1/2,0.9]$ we have $(\ln g_r)'' <
0$. Now,
\begin{eqnarray}
 (\ln g_r(\alpha))''
& = & r \left( \frac{f''(\alpha)}{f(\alpha)}
         - \frac{f'(\alpha)^2}{f(\alpha)^2} \right)
- \frac{1}{\alpha (1-\alpha)} \nonumber\\
& \leq & r \,\frac{f''(\alpha)}{f(\alpha)} - \frac{1}{\alpha
(1-\alpha)} \enspace . \label{eq:sec_der_b}
\end{eqnarray}
To show that the r.h.s.\ of~\eqref{eq:sec_der_b} is negative we
first note that for $\alpha \geq 1/2$ and $k > 3$,
\[ f''(\alpha) = 2^{1-k} k(k-1) (\alpha^{k-2} + (1-\alpha)^{k-2})
   < 2^{2-k} \alpha^{k-2} k^2 \]
is monotonically increasing. Therefore, $f''(\alpha) \leq f''(0.9)
< 2^{2-k} \,0.9^{k-2} \,k^2$.

Moreover, for all $\alpha$, $f(\alpha) \geq f(1/2)= (1-2^{-k})^2$.
Therefore, since $1/(\alpha (1-\alpha)) \geq 4$ and $r \leq
2^{k-1} \ln 2$, it suffices to observe that for all $k \ge 74$,
\[ (2^{k-1} \ln 2) \times \frac{2^{2-k} \,0.9^{k-2} \,k^2}{(1-2^{-74})^2}
 - 4 < 0 \enspace .
\]

Finally, recalling that $g'(1/2) = 0$ and using
\[
(\ln g_r)'' =    \frac{g''_r(\alpha)}{g_r(\alpha)}
         - \frac{g'_r(\alpha)^2}{g_r(\alpha)^2}
\]
we see that $g''_r(1/2)<0$ since $(\ln g_r)''(1/2)<0$.\hfill$\Box$

\bigskip
\noindent {\bf Proof of Lemma~\ref{lem:far}.}  By the definition
of $g_r$ we see that $g_r(\alpha) < g_r(1/2)$ if and only if
\begin{equation}
\label{neq:est}
        \left( \frac{f(\alpha)}{f(1/2)} \right)^r
    < 2 \alpha^{\alpha} (1-\alpha)^{1-\alpha} \enspace .
\end{equation}
Letting $h(\alpha) = - \alpha \ln \alpha - (1-\alpha) \ln
(1-\alpha)$ denote the entropy function, we see
that~\eqref{neq:est} holds as long as
\[ \frac{r}{\ln 2 - h(\alpha)} < \frac{1}{\ln (1+w)} \]
where
\[ w = \frac{f(\alpha) - f(1/2)}{f(1/2)} \enspace .
\]
Observe now that for $k > 3$, $f$ is strictly increasing in
$(1/2,1]$, so $w > 0$.  Moreover, for any $x>0$
\[
\frac{1}{\ln(1+x)} \geq \frac{1}{x}+\frac{1}{2}-\frac{x}{12}
\enspace .
\]
Since $f(\alpha)-f(1/2) = 2^{1-k} (\alpha^k+(1-\alpha)^k-2^{1-k})
< 2^{1-k}$ and $f(1/2) = (1-2^{1-k})^2 > 1 - 2^{2-k}$, we thus see
that it suffices to have
\begin{equation}
\label{eq:need}
  \frac{r}{\ln 2 - h(\alpha)}
< \frac{2^{k-1} - 2}{\alpha^k + (1-\alpha)^k - 2^{1-k}} +
\frac{1}{2} - \frac{2^{1-k}}{12} \enspace .
\end{equation}

Now observe that for any $0 < \alpha < 1$ and $0 \le q <
\alpha^k$,
\[ \frac{1}{\alpha^k - q}  \ge  1 + k(1-\alpha) + q \enspace . \]
Since $\alpha > 1/2$ we can set $q = 2^{1-k} - (1-\alpha)^k$,
yielding
\[
\frac{1}{\alpha^k + (1-\alpha)^k - 2^{1-k}} \,\ge\, 1 +
k(1-\alpha) + 2^{1-k} - (1-\alpha)^k \enspace .
\]
Since $2^k (1-\alpha)^k < 5^{-k}$, we find that~\eqref{neq:est}
holds as long as $r \leq \phi(y) - 2^{-k}$ where
\[
 \phi(\alpha) \equiv \bigl(\ln 2 - h(\alpha) \bigr)
 \!\! \left( 2^{k-1} + (2^{k-1}-2) k (1-\alpha) - \frac{1}{2} \right)
  .
\]

We are thus left to minimize $\phi$ in $(0.9,1]$. Since $\phi$ is
differentiable its minima can only occur at $0.9$ or $1$, or where
$\phi'=0$. The derivative of $\phi$ is
\begin{equation}
\phi'(\alpha) =  (2^{k-1} - 2) \times \Biggl[ -k \,(\ln 2 -
h(\alpha)) + (\ln \alpha - \ln (1-\alpha)) \left( 1 + k(1-\alpha)
+ \frac{3}{2^k-4} \right) \Biggr] \enspace . \nonumber
\label{eq:phiprime}
\end{equation}
Note now that for all $k>1$
\[ \lim_{\alpha \to 1} \phi'(\alpha) = - \frac{2^{k}-1}{2} \ln (1-\alpha) \]
is positively infinite.  At the same time,
\[ \phi'(0.9) < -0.07 \times 2^k k + 1.1 \,(2^k-1) + 0.3 \,k \]
is negative for $k \ge 16$.    Therefore, $\phi$ is minimized in
the interior of $(0.9,1]$ for all $k \geq 16$. Setting $\phi'$ to
zero gives
\begin{equation}
\label{eq:boot} - \ln (1-\alpha)
 = \frac{k \,(\ln 2 - h(\alpha))}
        {1 + k (1-\alpha) + 3/(2^k-4)}
 - \ln \alpha \enspace .
\end{equation}

By ``bootstrapping'' we derive a tightening series of lower bounds
on the solution for the l.h.s.\ of~\eqref{eq:boot} for $\alpha \in
(0.9,1)$.  Note first that we have an easy upper bound,
\begin{equation}
\label{eq:phiupper} - \ln (1-\alpha) < k \ln 2 - \ln \alpha
\enspace .
\end{equation}
At the same time, if $k > 2$ then $3/(2^k-4) < 1$, implying
\begin{equation}
\label{eq:philower}
 - \ln (1-\alpha)
 > \frac{k \,(\ln 2 - h(\alpha))}
        {2 + k (1-\alpha)}
 - \ln \alpha
\enspace .
\end{equation}
If we write $k(1-\alpha)=B$ then~\eqref{eq:philower} becomes
\begin{equation}\label{eq:contra}
- \ln (1-\alpha)
 >  \frac{\ln 2 - h(\alpha)}{1-\alpha}
  \left( \frac{B}{B+2} \right)
  - \ln \alpha \enspace .
\end{equation}

By inspection, if $B \geq 3$ the r.h.s.\ of~\eqref{eq:contra} is
greater than the l.h.s.\ for all $\alpha > 0.9$, yielding a
contradiction. Therefore, $k(1-\alpha) < 3$ for all $k > 2$. Since
$\ln 2 - h(\alpha) > 0.36$ for $\alpha > 0.9$, we see that for
$k>2$, \eqref{eq:philower} implies
\begin{equation}\label{eq:rissoto}
- \ln (1-\alpha) > 0.07 \,k \enspace .
\end{equation}
Finally, observe that~\eqref{eq:rissoto} implies that as $k$
increases the denominator of~\eqref{eq:boot} approaches $1$.

To bootstrap, we note that since $\alpha > 1/2$ we have
\begin{eqnarray}
 h(\alpha) & \le & -2 (1-\alpha) \ln (1-\alpha) \label{eq:ent_b_l}\\
           & < & 2 \,\e^{-0.07 \,k} (k \ln 2 - \ln 0.9) \label{eq:randkl} \\
           & < & 2 \,k \,\e^{-0.07 \,k} \nonumber
\end{eqnarray}
where~\eqref{eq:randkl} relies on~\eqref{eq:phiupper}
and \eqref{eq:rissoto}. Moreover, $\alpha > 1/2$
implies $-\ln \alpha \le 2(1-\alpha) < 2 \,\e^{-0.07 \,k}$.
Thus, by using~\eqref{eq:rissoto} and the fact $1/(1+x)>1-x$ for
all $x > 0$, \eqref{eq:boot} gives for $k \geq 3$,
\begin{eqnarray}
 -\ln (1-\alpha) & > & \frac{k \,(\ln 2 - h(\alpha))}
                            {1 + k (1-\alpha) + 3/(2^k-4)} \nonumber \\
  & > & \frac{k \,(\ln 2 - 2 \,k \,\e^{-0.07 \,k}) }
             {1 + 2 \,k \,\e^{-0.07 \,k}} \nonumber \\
  & > & k \,(\ln 2 - 2 \,k\,\e^{-0.07 \,k}) (1 - 2 \,k \,\e^{-0.07 \,k} )
\nonumber \\
  & > & k \ln 2 - 4 \,k^2\,\e^{-0.07 \,k} \enspace . \label{eq:fried}
\end{eqnarray}
For $k \ge 166$, $4 \,k^2 \,\e^{-0.07 \,k} < 1$. Thus, by
\eqref{eq:fried}, we have $1 - \alpha < 3 \times 2^{-k}$. This, in
turn, implies $-\ln \alpha \leq 2(1-\alpha) < 6 \times 2^{-k}$ and
so, by \eqref{eq:ent_b_l} and~\eqref{eq:phiupper}, we have for
$\alpha > 0.9$
\begin{equation}\label{eq:entr_last}
h(\alpha) < 6 \times 2^{-k} (k \ln 2 - \ln \alpha)
          < 5 \,k \,2^{-k} \enspace .
\end{equation}

Plugging~\eqref{eq:entr_last} into~\eqref{eq:boot} to bootstrap
again, we get that for $k \geq 3$
\begin{eqnarray*}
 -\ln (1-\alpha) & > & \frac{k \,(\ln 2 - 5 \,k \,2^{-k})}
                            {1 + 3 \,k\,2^{-k} + 3/(2^k-4)} \\
 & > &  \frac{k \,(\ln 2 - 5 \,k \,2^{-k})}
             {1 + 6 \,k \,2^{-k}}\\
 & > & k \,(\ln 2 - 5 \,k \,2^{-k}) (1 - 6 \,k \,2^{-k})\\
  & > & k \ln 2 - 11 \,k^2 \,2^{-k}  \enspace .
\end{eqnarray*}
Since $e^x < 1+2x$ for $x < 1$ and $11 \,k^2 \,2^{-k} < 1$ for $k
> 10$, we see that for such $k$
\[ 1-\alpha < 2^{-k} + 22 \,k^2 \,2^{-2k} \enspace .
\]

Plugging into~\eqref{eq:phiupper} the fact $-\ln \alpha < 6 \times
2^{-k}$ we get $-\ln (1-\alpha) < k \ln 2 + 6 \times 2^{-k}$.
Using that $\e^{-x} \geq 1-x$ for $x\geq 0$, we get the closely
matching upper bound,
\[ 1-\alpha > 2^{-k} - 6 \times 2^{-2k} \enspace . \]

Thus, we see that for $k \ge 166$, $\phi$ is minimized at an
$\alpha_{\min}$ which is within $\delta$ of $1-2^{-k}$, where
$\delta = 22 \,k^2 \,2^{-2k}$. Let $T$ be the interval
$[1-2^{-k}-\delta, 1-2^{-k}+\delta]$. Clearly the minimum of
$\phi$ is at least $\phi(1-2^{-k}) - \delta \times \max_{\alpha
\in T} |\phi'(\alpha)|$. It is easy to see
from~\eqref{eq:phiprime} that if $\alpha \in T$ then
$|\phi'(\alpha)| \le 2 \,k \,2^k$.

Now, a simple calculation using that $\ln (1-2^{-k}) > - 2^{-k} -
2^{-2k}$ for $k \ge 1$ gives
\begin{eqnarray*}
 \phi(1-2^{-k})
& = & \frac{1}{2}
      \bigl( (2^k - k) \ln 2 + (2^k - 1) \ln (1-2^{-k}) \bigr)  \, \times \,\bigl( 1 + (k-1) \,2^{-k} - k \,2^{2-2k} \bigr) \\
& > & 2^{k-1} \ln 2 - \frac{\ln 2}{2} - \frac{1}{2} - k^2 \,2^{-k}
\enspace .
\end{eqnarray*}
Therefore,
\[ \phi_{\min}
 \ge 2^{k-1} \ln 2 - \frac{\ln 2}{2} - \frac{1}{2} - 45 \,k^3 \,2^{-k}
\enspace .
\]
Finally, recall that~\eqref{neq:est} holds as long as $r <
\phi_{\min} - 2^{-k}$, \ie
\[ r
 < 2^{k-1} \ln 2 - \frac{\ln 2}{2} - \frac{1}{2} - 46 \,k^3 \,2^{-k}
\enspace .
\]
Clearly, we can take $k_0=O(\ln \eps^{-1})$ so that for all $k
\geq k_0$ the error term $46 \,k^3 \,2^{-k}$ is smaller than any
$\eps > 0$.  \mbox{}\hfill$\Box$

\section{Proof of Lemma~\ref{lem:peak}}
\label{sec:peak}

The idea behind 
Lemma~\ref{lem:peak} 
is that sums of this type are dominated by the contribution of
$\Theta(n^{1/2})$ terms around the maximum term. The proof amounts
to replacing the sum by an integral and using the Laplace method
for asymptotic integrals~\cite{debruijn}.

We start by establishing two upper bounds for the
terms of $S_n$, 
one crude and one sharp. For the sharp bound we will use the
following form of Stirling's approximation, valid for all $n > 0$:
\begin{equation}
\label{eq:stirling} \sqrt{2 \pi n}
 < \frac{n!}{(n/\e)^{n}}
 < \sqrt{2 \pi n} \,\left( 1+1/n\right)
   \enspace .
\end{equation}

Recall that the $z$th term of $S_n$ is $\binom{t}{z}^{\!q}
\,\phi(z/t)^n$, where $n=qt$ and $\phi(\alpha) = g(\alpha) \,
\alpha^\alpha (1-\alpha)^{1-\alpha}$. Fix any $\delta>0$ and
suppose that $z = \alpha t$ where $\alpha \in [\delta,1-\delta]$.
Then~\eqref{eq:stirling} yields
\begin{equation}\label{eq:binbon}
{\binom{t}{z}^{\!q} \,\phi(z/t)^n} < {
  \,s(\alpha) \,g(\alpha)^n} \left(1 + \frac{q}{n}
\right)^q \enspace ,
\end{equation}
where $s(\alpha) = \left(2\pi\alpha(1-\alpha)t\right)^{-q/2}$. In
addition to~\eqref{eq:binbon}, valid for $z \in [t\delta,
t(1-\delta)]$, we will also use a cruder bound, valid for all $0
\le z \le t$. Namely, using the upper bound of~\eqref{eq:stirling}
for $t!$ and the lower bound $n! > (n/\e)^{n}$ for $z!$ and
$(t-z)!$ (where we take $0^0 \equiv 1$) we get (since $1+1/t \leq 2$)
\begin{equation}
\label{eq:bincrude} \binom{t}{z}^{\!q} \,\phi(z/t)^n \;<\; \left(
{8 \pi n}\right)^{q/2}
   \,g(\alpha)^n
\enspace .
\end{equation}

Recall now that $g(\amax)> g(\alpha)$ for all $\alpha \neq \amax$.
If $I_\eps$ denotes the interval $[\amax-\eps, \amax+\eps]$ then
for every $\eps > 0$, there exists a constant $g_{\eps} <
g(\amax)=\gmax$ such that $g(\alpha) < g_{\eps}$ for all $\alpha
\notin I_\eps$. Let $z^-_{\eps} = \lfloor (\amax-\eps) t \rfloor$
and $z^+_{\eps} = \lceil (\amax+\eps) t \rceil$, and let
\begin{equation}
\label{eq:sepsdef}
S^{(\eps)}_n =  \sum_{z = z^-_{\eps}}^{z^+_{\eps}}
  \binom{t}{z}^{\!q} \,\phi(z/t)^n \enspace .
\end{equation}
We use~\eqref{eq:binbon} to bound the terms in $S^{(\eps)}_n$
and~\eqref{eq:bincrude} to bound the remaining terms of $S_n$.
Since $\lim_{n \to \infty} \left(1 + q/{n} \right)^q= 1$, and
since $\lim_{n \to \infty} n^s \,g_{\eps}^n / \gmax^n = 0$ for any
$s$, we see that for every $\eps> 0$
\begin{equation}
\label{eq:sequalsseps} S_n < 
(2 \pi t)^{-q/2} \times \sum_{z = z^-_{\eps}}^{z^+_{\eps}}
g(z/t)^n \enspace .
\end{equation}

Say that a twice-differentiable function $\psi(x)$ is {\em
unimodal} on an interval $[a,b]$ if $\psi'$ has a unique zero $c
\in [a,b]$ with $a < c < b$, and furthermore $\psi''(c) < 0$.
Since $\gmax>g(\alpha)$ for all $\alpha > \amax$ and $g''(\amax)
<0$, we can take $\eps$ small enough so that $g$ is unimodal on
$I_\eps$. This implies that $\ln g$ is also unimodal on $I_{\eps}$
and, for $n \geq 1$, that $g^n$ is unimodal also. Since $g^n$ is
unimodal,
\begin{equation}
\label{eq:sumintegral} \sum_{z = z^-_{\eps}}^{z^+_{\eps}} g(z/t)^n
\leq n \int_{I_{\epsilon}} g(x)^n \,\dx \,+ \gmax^n\enspace .
\end{equation}
We evaluate this last integral using Lemma~\ref{lem:debruijn}, \ie
the Laplace method for asymptotic integrals.
\begin{lemma}\label{lem:debruijn}{\cite[\S 4.2]{debruijn}.} Let
$h(x)$ be unimodal on $[a,b]$ where $c$ is the unique zero of $h'$
in $[a,b]$.  Then
\[ \lim_{n \to \infty} \int_a^b \e^{nh(x)} \,\dx
 \, = \, \sqrt{\frac{2 \pi}{n \,|h''(c)|}} \,\e^{nh(c)}
\]
\end{lemma}
Applying Lemma~\ref{lem:debruijn} to \eqref{eq:sumintegral} with
$h= \ln g$ and $c = \amax$, we see that $S_n < C \, n^{-(q-1)/2}
\,\gmax^n$, where $C= (2\pi)^{-(q-1)/2} \times q^{q/2} \times
\sqrt{\gmax/g''(\amax)}$.\hfill$\Box$

\bigskip
Lemma~\ref{lem:peak} has the following obvious corollary, 
which is useful for a variety of second moment calculations.
\begin{corollary}
Let $S_n$ and $g(\alpha)$ be defined as in Lemma~\ref{lem:peak}.  
If there exists $\amax \in (0,1)$ with $g(\amax) \equiv \gmax$ 
and a constant $A > 0$ such that $g(\alpha) \le \gmax - A (\alpha-\amax)^2$ 
for all $\alpha$, then there exists a constant $C = C(q,\gmax,\amax,A) > 0$ 
such that
\[ 
S_n < C \,{n^{-(q-1)/2} \,\gmax^n} \enspace .
\]
\end{corollary}

\comment{
\section{Upper bounds for NAE $k$-SAT and Hypergraph 2-colorability}
\label{sec:upper}

As mentioned in the Introduction, our upper bounds for the
hypergraph 2-colorability threshold and for the NAE $k$-SAT
threshold  follow by a direct application of the technique
of~\cite{KKKS,dubo}. For the sake of completeness we carry out the
proof below for NAE $k$-SAT. The proof for hypergraph
2-colorability is essentially identical.

Let us say that a NAE-satisfying assignment $\sigma$ is {\em
locally maximal\/} if no 0 in $\sigma$ can be switched to 1 while
maintaining NAE-satisfiability. For any formula $F$ let $X(F)$ be
the number of locally maximal NAE-satisfying assignments. Observe
that $X=X(F)>0$ iff $F$ is NAE-satisfiable and, therefore, proving
$\ex[X] = o(1)$ implies NAE-unsatisfiability \whp\ To compute
$\ex[X]$ this time it will be convenient to work with random
formulas on $n$ variables with $m=rn$ clauses where each clause is
selected with replacement from the set of all $2^k \binom{n}{k}$
proper clauses (see Section~\ref{sec:gene}).

We start by observing that if $\sigma$ NAE-satisfies a formula
$F$,  then a variable $v$ assigned 0 by $\sigma$ cannot be
switched to 1 iff $F$ contains at least one $v$-blocking clause,
\ie a clause $c$ in which $v$'s literal is either the only
satisfied or the only unsatisfied literal $c$. Moreover, observe
that for every $\sigma$ and $v$ the number of potential
$v$-blocking clauses with respect to $\sigma$ is
$2\binom{n-1}{k-1}$ and that these sets of clauses are disjoint
for distinct variables. Therefore, if $\sigma$ is any assignment
assigning 0 to variable $v$ and $F$ is a random $k$-CNF formula
with $m$ clauses, then conditional on $\sigma$ being
NAE-satisfying, the probability that $v$ is blocked is precisely
\[
b = 1 -   \left(1 -
\frac{2\binom{n-1}{k-1}}{(2^k-2)\binom{n}{k}}\right)^m
    = 1 -   \left(1 -
\frac{2k}{(2^k-2)n}\right)^m = 1 -
\exp\left(-\frac{2kr}{2^k-2}\right) + O(1/n) \enspace .
\]

Since the sets of blocking clauses for different variables are
disjoint and since the number of clauses is fixed, the events
corresponding to the blocking of distinct variables are negatively
correlated (see \eg\cite{mcd}). As a result, if the number of 0s
in $\sigma$ is $z$, the probability that $\sigma$ is a locally
maximal NAE-satisfying assignment is at most
\[
 \left(1- 2^{1-k}\right)^{m} b^z \enspace .
\]
Using the binomial theorem to sum over all truth assignments, \ie over $z \in
\{0,\ldots,n\}$, we get
\[
\ex[X] \leq \left(1- 2^{1-k}\right)^{m} \left(1+b\right)^n
\enspace .
\]
Thus, a constant $r$ is an upper bound on the NAE $k$-SAT
threshold whenever
\begin{equation}\label{banff}
\left(1- 2^{1-k}\right)^{r} \left(2-\exp(-2kr/(2^k-2))\right) < 1
\enspace .
\end{equation}
Inequality~\eqref{banff} holds for the values of $r$ presented in
Table~\ref{tab:val} and, asymptotically, for
\[
{r = 2^{k-1}\ln2 - \frac{\ln 2}{2} - \frac{1}{4} + O({k}{2^{-k}})}
\]
completing the proof of Theorem~\ref{thm:naeksat}.
\hfill$\Box$
}

\section{Conclusions}\label{sec:conc}

Before this work, lower bounds on the thresholds of random
constraint satisfaction problems were largely derived by analyzing
very simple heuristics. Here, instead, we derive such bounds by
applying the second moment method to the number of solutions. In
particular, for random NAE $k$-SAT and random hypergraph
2-colorability we determine the location of the threshold within a
small additive constant for all $k$. As a corollary, we establish
that the asymptotic order of the random $k$-SAT threshold is
$\Theta(2^k)$ answering a long-standing open question.

Since this work first appeared~\cite{focs,usrandom}, our methods
have been extended and applied to other problems. For random
$k$-SAT, Achlioptas and Peres~\cite{yuval} confirmed our suspicion
(see Section~\ref{sec:boost}) that the main source of correlations
in random $k$-SAT is the ``populist'' tendency of satisfying
assignments towards the majority vote assignment. By considering a
carefully constructed random variable which focuses on balanced
solutions, \ie on satisfying assignments that satisfy roughly half
of  all literal occurrences, they showed $r_k \geq 2^k \ln 2 - k/2
-O(1)$, establishing $r_k \sim 2^k \ln 2$.

In~\cite{max}, Achlioptas, Naor and Peres extended the approach of
balanced solutions to Max $k$-SAT. Let us say that a $k$-CNF
formula is $p$-satisfiable if there exists a truth assignment
which satisfies at least $(1-2^{-k}+p2^{-k})$ of all clauses; note
that every $k$-CNF is 0-satisfiable. For $p \in (0,1]$ let
$r_k(p)$ denote the threshold for $F_k(n,m=rn)$ to be
$p$-satisfiable (so that $r_k(1) = r_k$). In~\cite{max}, the
result $r_k =r_k(1) \sim 2^k \ln 2$ of~\cite{yuval} was extended
to all $p \in (0,1]$ showing
\[
r_k(p) \sim  \frac{2^k \ln 2}{p+(1-p)\ln(1-p)} \enspace .
\]

In both~\cite{yuval} and~\cite{max}, controlling the variance
crucially depends on focusing on an appropriate subset of
solutions (akin to our NAE-assignments, but less heavy-handed).
In~\cite{birk}, Achlioptas and Naor applied the naive second
moment method to the canonical symmetric constraint satisfaction
problem, \ie to the number of $k$-colorings of a random graph.
Bearing out our belief that the naive approach should work for
symmetric problems they obtained asymptotically tight bounds for
the $k$-colorability threshold. The difficulty there is that the
``overlap parameter" is a $k \times k$ matrix rather than a single
real $\alpha \in [0,1]$. Since $k \to \infty$, this makes the
asymptotic analysis dramatically harder and much closer to the
realm of  statistical mechanics calculations.\medskip

We propose several questions for further work.
\begin{enumerate}
\item Does the second moment method give tight lower bounds on
the threshold of all constraint satisfaction problem with a permutation symmetry?
\item Does it perform  well for problems that are symmetric ``on average"?
For example, does it perform well for {\em regular\/} random
$k$-SAT where every literal appears an equal number of times?
\item What rigorous connections can be made between the success of the
second moment method and the notion of ``replica symmetry'' in
statistical physics?
\item 
Is there a polynomial-time algorithm that succeeds with uniformly
positive probability close to the threshold, or at least for $r =
\omega(k) \times  2^k/k$ where $\omega(k) \to \infty$?
\end{enumerate}

\medskip
\paragraph{Acknowledgements} We are grateful to Paul Beame, Ehud Friedgut,
Michael Molloy, Assaf Naor, Yuval Peres, Alistair Sinclair, and Chris Umans for reading
earlier versions and making many helpful suggestions, and to Remi
Monasson for discussions on the replica method. We would like to
thank Henry Cohn for bringing~\cite{debruijn} to our attention.
C.M. is funded partly by the National Science
Foundation under grant PHY-0200909, and thanks Tracy Conrad for her support.

\def\bibRCS{$Id: theory.bib,v 1.359 2000/09/08 01:00:19 beame Exp beame $}
  \makeatletter \@ifundefined{ccisdefined}{ \newcommand{\cc}[1]{\mbox{\it
  #1\/}} \newcommand{\ccisdefined}{} }{} \@ifundefined{journalfont}{
  \newcommand{\journalfont}{\it } }{} \makeatother \def\bibRCS{$Id:
  theory.bib,v 1.366 2000/09/15 17:54:09 beame Exp $} \makeatletter
  \@ifundefined{ccisdefined}{ \newcommand{\cc}[1]{\mbox{\it #1\/}}
  \newcommand{\ccisdefined}{} }{} \@ifundefined{journalfont}{
  \newcommand{\journalfont}{\it } }{} \makeatother

\end{document}